# An analysis of astronomical alignments of Greek Sicilian Temples

Alun Salt, The Centre for Interdisciplinary Science, University of Leicester,

e: alun.m.salt@gmail.com t: @alun

## Abstract

In the eighth century BC something peculiar seems to happen on Sicily. The archaeological record starts to show the arrival of Greek material culture. By the fifth century BC the island is effectively 'Hellenised' and ancient historians record the political and military action of poleis, Greek city-states. Each polis has traditionally been seen as the offshoot of a city elsewhere. Genealogies of cities ultimately end in cities found in the cities of the Peloponnese and the Aegean. The 'Greek' identity of the Sicilian cities is part of a wider debate on the concept of Identity in the ancient world. This paper considers if there is a contribution archaeoastronomers can make to such discussions by considering the alignments of Greek temples. Greek religion was intimately related to notions of civic identity and what it meant to be 'Greek'. I propose a method of studying small samples of temples, which combines both alignment analysis and historical context. Therefore it may be possible that a study of the temples may yield useful information about collective identities. However, as this method shows, the more ambiguous the cultural data the less certain any astronomical patterns may be.

## Binomial Distribution and the temples of Greek Sicily

This paper develops a method initially published by Salt (2009) using data published in the same paper. The data set comprises 41 orientations of 44 temples dating from the archaic and classical periods found in ten poleis around Sicily. Some temples have been combined as one data point. One example would be the Temple A/B at Himera. The Temple B overlies the Temple A, being built shortly after the smaller temple (Bonacasa, 1970). The reason for combining the two temples is that the alignment of Temple B seems to be defined by the alignment of Temple A. Therefore the two temples may not be independently aligned. A response to this approach could be to argue that the Greeks were perfectly capable of building over older temples with differing alignments as the ruins at Gela show (Orlandini, 1968). Therefore a reason the two temples share an alignment is because the alignment is important rather than a happy accident of architecture. I have chosen to continue the practice of combining of possibly co-planned temples. Temples A and O and F and G may be pairs of temples planned along a shared axis, and so may not have been independently aligned. When showing an intentionally astronomical motivation for the orientation of a temple, this method will yield more conservative results, which is desirable when presenting results to a sceptical reader.

The method in the 2009 paper was an extremely simple application of the Binomial Distribution to the answer the question of whether or not Greek temples pointed 'east'. The advantages of this method are its simplicity, its emphasis on the importance of cultural information in defining a test probability and its applicability to small samples. The output is related to the size of the sample and so, unlike some statistical tests, smaller samples will produce less emphatic results. The binomial distribution is given by the formula

$$P(k;n;p) = \binom{n}{k} p^k (1-p)^{41-k}$$

where $n$ is the sample size, $p$ is the probability an event will occur and $k$ is the number of occurrences. This is not a friendly formula. The key feature of the binomial distribution can be calculated by two considerably simpler formulae. The peak of the distribution can be calculated by the formula

$$\text{peak} = np$$

where $n$ is the sample size and $p$ is the probability between 0 and 1. The standard deviation of the distribution, σ is calculated by the formula

$$\sigma = \sqrt{(np(1-p))}$$

Therefore the expected probability of temples facing the eastern half of the horizon in sample size of is 41x0.5, or 20.5. The actual result, 40, lies six standard deviations from the peak value.

There is no formula for significance itself. This is for the reader to judge. Typically social scientists will attempt to demonstrate an effect to 2σ, roughly a 95% confidence interval. It should be noted that these effects are supported by a variety of cultural factors, which give the experimenter confidence that the test is meaningful. In contrast Schaeffer (2006a, p. 29) notes in astronomical research that 3σ claims are wrong 50% of the time and that a target of 4σ or 5σ would be preferable. The reason for the difference is that astronomical data has no social context and so we cannot always be sure we are comparing culturally coherent samples. Expressing results in terms of standard deviations away from the predicted result allows the reader to determine whether or not a claim demonstrates significance.

In the case of the 2009 paper, the results vary according to what definition of 'east' is used. If 'east' means the range of horizon marked by the rising sun the number of temples facing east, 38, is 8.3σ away from the expected result of 6.8 temples. The 2009 paper goes no further than analysing the orientations of Sicilian temples as a whole. Is it possible to isolate sub-samples and see if significant distributions of alignments occur at local scales?

I shall start by considered the temples of Akragas. Is there something specifically Akragan about the choices of temple alignments? If the inhabitants of Akragas had a preference for one alignment, 90º to take an example, then I could examine the temples at Akragas to see if they were more likely to face 90º than typical Greek Sicilian temples. This sadly is not the case. The temples face many directions, but they could still be clustered in atypical targets. Working out if this creates an unusual skew in the temple dataset is not straightforward.

What are the odds that a temple would face the sort of direction that the Akragans chose to orientate their temples to? Of the 41 data points in the Sicilian set, 9 obviously face this kind of direction – the nine temples found at Akragas. Additionally, the temple of Zeus Melikhios at Selinous faces 80º, which is the orientation of the temples of Zeus Olympios and Temple I at Akragas. So the probability of an Akragan alignment would appear to be 10/41 or 0.244. Are there more Akragan alignments than we would expect in the Sicilian data set? We would expect 41x0.244 temples in a set of 41, or 10. This method clearly does not work. All it shows is that the temples that face the direction of Akragan temples are the temples that face those directions.

The reason this method cannot work is that when we are calculating a value for *p* we are not calculating the probability that a typical Greek Sicilian temple is faces in an Akragan direction. We are calculating the probability of a set which includes typical Greek Sicilian temples *and* the temples from Akragas which may or may not be similar to typical temples. To calculate the probability that typical Greek Sicilian temples face an Akragan direction it necessary to exclude the Akragan temples from the initial calculation.

The alignments of the temples at Akragas give us the list of directions an Akragan temple would face: 80º (twice), 81º 82º, 87º (twice), 90º (twice), 110º. If this reflects a cycle of ritual behaviours then this is not a continuous range, but rather fragmentary. Therefore it is a matter of which ordinary Greek Sicilian temples face these events.

I shall now borrow an assertion from McCluskey who had a similar problem analysing the alignments of medieval churches in England. McCluskey (2006, p. 415) has argued that churches aligned to specific saints would not be intentionally orientated to avoid certain azimuths. Likewise I shall use the assumption that the alignments of Sicilian temples may indicate a preference for certain alignments, but not away from specific orientations. Therefore if I examine any specific sub-group of temples, the remainder of temples from the data set form a usable comparison sample set of typical Greek Sicilian temples.

Of these remaining 32 temples just one of them, the temple of Zeus Melikhios at Selinous, faces in one of these directions. Therefore the probability that a typical Greek Sicilian temple will face in an Akragan direction is 1 in 32. In terms of the formulae above $n$ is 32 and $p$ is 0.03 with the standard deviation $\sigma$ of a binomial distribution being 0.98.

The sample set of typical Greek Sicilian temples cannot be directly compared to the data set I observed. My set has 41 temples, the set of typical temples is too small. To increase the size of the typical set to 41 temples for comparison I must multiply $n$ by 41/32 or 1.28. Now for a set of 41 typical Greek Sicilian temples we would expect $np$ or 41x0.03 temples to face the same direction as an Akragan temple. The standard deviation of a binomial distribution would now be 1.11. Obviously the data set including the Akragan temples will have more alignments. The question is, are there so many more Akragan temples that the number of alignments is *significantly* more? There are 10 Akragan alignments in the data set, 8.72 temples more than the expected value. This is 8.72/1.11 or 7.83 standard deviations away from the expected value. If you are satisfied with claims 5$\sigma$ or higher, this would be a significant result. If you prefer claims to be 10$\sigma$ or higher then there is nothing special about temples at Akragas.

There would be a few problems with this approach.

Firstly there is the matter of the accuracy of the data. The data is recorded to an accuracy of ±1º. Therefore any temple which is has an alignment a degree different to any temple at Akragas should also be considered a match. This produces many more matches. The Temple B at Gela faces an azimuth of 111º, while the Temple E at Akragas faces 110º. Allowing for errors, these temples may be facing the same direction. Similarly, the Heroon and Temple Sud-Est at Megara Hyblaea and the temple of Demeter Malaphoros at Selinous all have orientations within a degree of temples at Akragas. Therefore the probability of a random Greek Sicilian temple having an Akragas-like alignment is in fact 5/32. If we apply this value of $p$ to a sample of 41 typical Greek

Sicilian temples we would expect to have 6.4 Akragan alignments with a standard deviation of 2.32. This means that the observed result of 14 temples having Akragas-like alignments in Sicily is only 3.27σ. The probability of this happening is less than one in a hundred and so I would argue is worth discussing, but nothing like as striking as the first figure.

A second objection is that I am also assuming that the Greeks themselves built temples to an accuracy of one degree. This is not an assumption the reader may share. Henriksson and Blomberg (2000, p. 304) have argued that the accuracy of astronomy in the *Phaenomena* of Aratus is two and a half degrees. Aratus was writing during the 3rd century BC, which would make a claim for 1º accuracy in earlier periods unlikely. While I do not agree with Henriksson and Blomberg for reasons I shall give below, I can accept that the reader may. Therefore it would be helpful if I could calculate what the probability is that temples would be thought of as facing the same direction if this level of accuracy were used.

If we count any temple with an alignment within two degrees of an Akragas alignment as a match, then the Temple Ouest at Megara Hyblaea and the Athenaion at Syracuse also match temples at Akragas. The result of 16 temples is now only 2.66σ from the expected result. At three degrees there are three further matches, the temple at Camarina, the Temple Sud a Colonnade at Megara Hyblaea, and Naxos's Temple C. Significance is now reduced to 2.08σ.

Even this may be too accurate for some readers, who will note that the Greeks had a lunar calendar and the same conceptual day of the year cover quite a range of solar declinations. Therefore the results below are calculated for assumed accuracies between 1º and 10º for solar azimuths.

Finally there is the matter that this method always produces a positive result. This test alone cannot define a sample as inherently significant. It can provide a guide as to whether or not a sample is worth investigating further and for comparing differences between samples. A sample that produces a staggeringly high result might seem purely statistically significant, but this significance would still be due to the value of $p$ and this figure must be justified with reference to some cultural information.

The calculations for Akragas produce the following results.

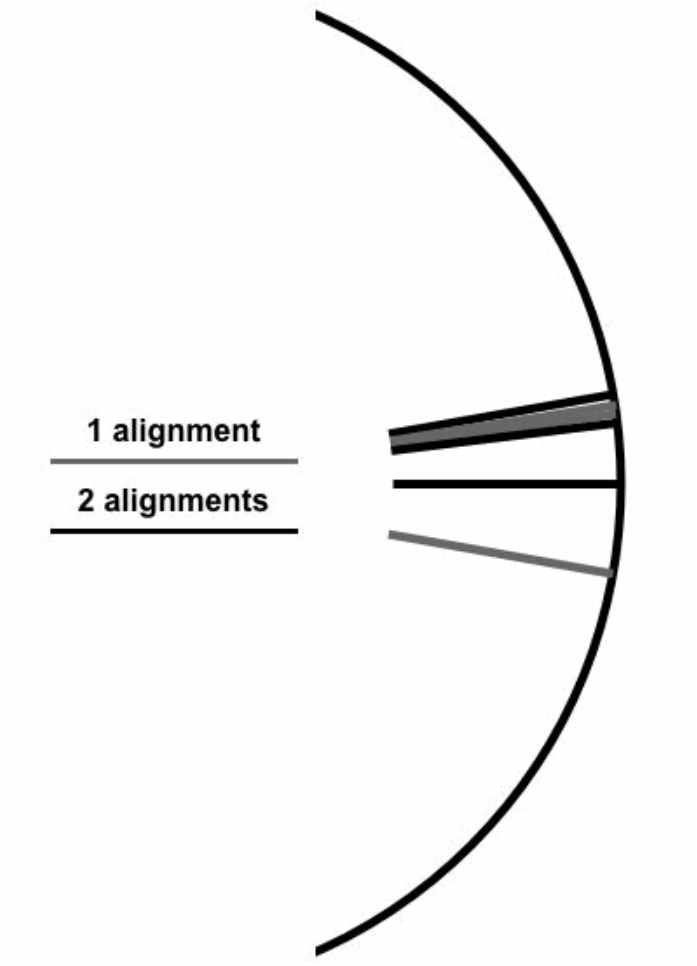

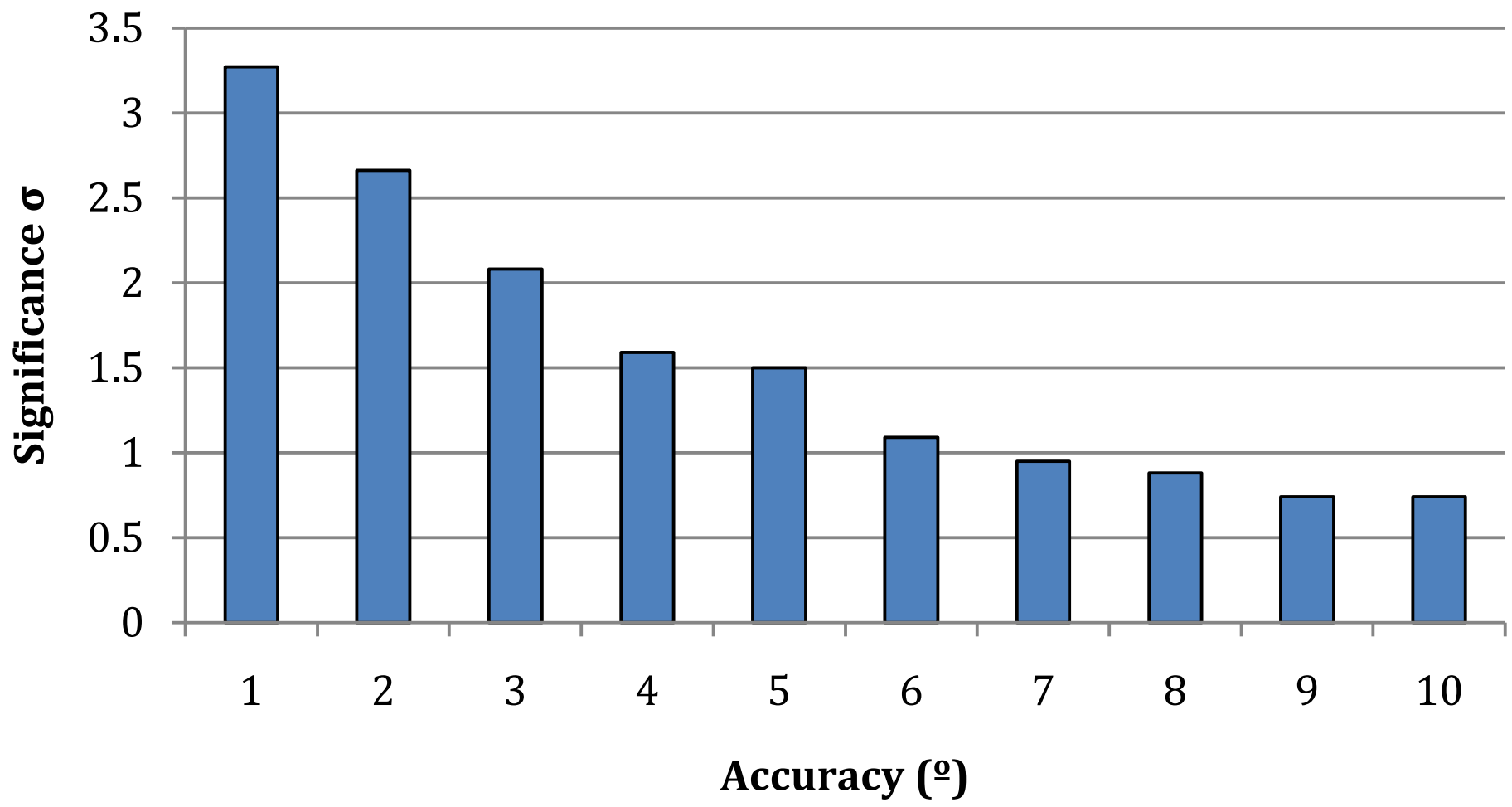

| Presumed Accuracy | 1º | 2º | 3º | 4º | 5º | 6º | 7º | 8º | 9º | 10º |
|---|---|---|---|---|---|---|---|---|---|---|
| Number of matches | 5 | 7 | 10 | 14 | 15 | 20 | 22 | 23 | 25 | 25 |
| Distance of result from expected value (σ) | 3.27 | 2.66 | 2.08 | 1.59 | 1.50 | 1.09 | 0.95 | 0.88 | 0.74 | 0.74 |

Similar tables and graphs have been produced for the azimuths of the other Greek poleis with three temples or more.

## Analysis of Greek temple alignments by location

### Gela

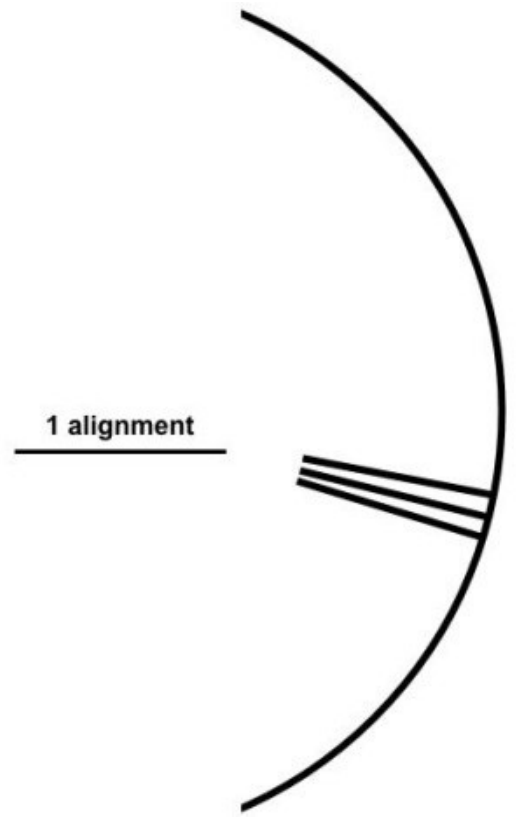

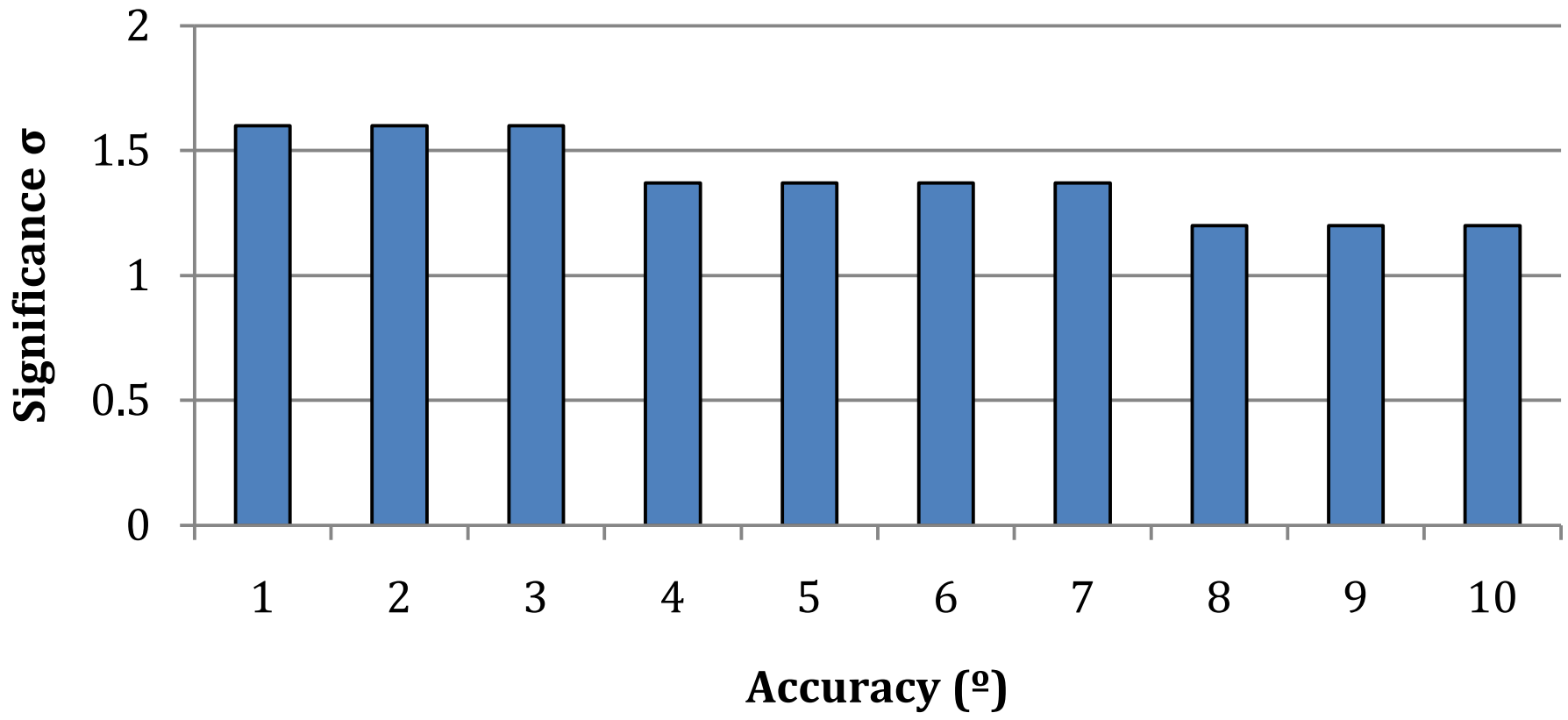

| Presumed Accuracy | 1º | 2º | 3º | 4º | 5º | 6º | 7º | 8º | 9º | 10º |
|---|---|---|---|---|---|---|---|---|---|---|
| Number of matches | 3 | 3 | 3 | 4 | 4 | 4 | 4 | 5 | 5 | 5 |
| Distance of result from expected value (σ) | 1.60 | 1.60 | 1.60 | 1.37 | 1.37 | 1.37 | 1.37 | 1.20 | 1.20 | 1.20 |

### Himera

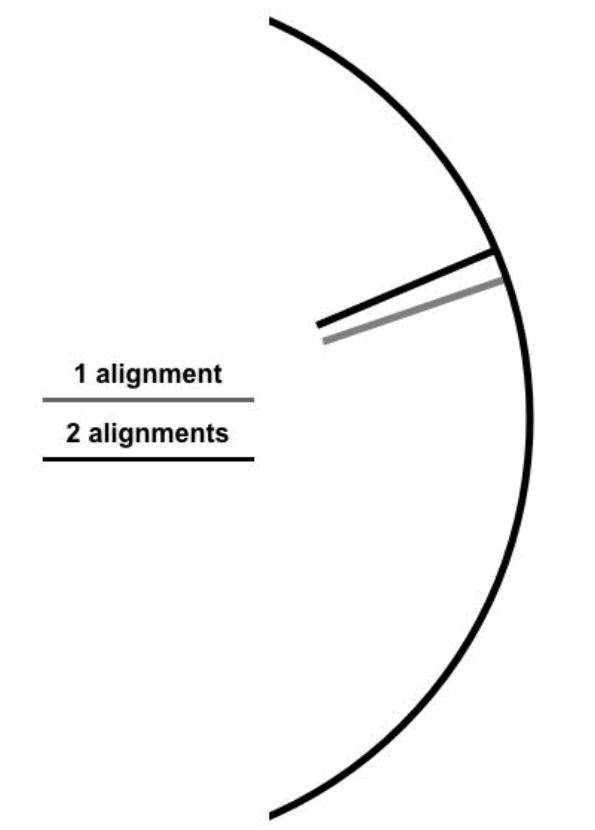

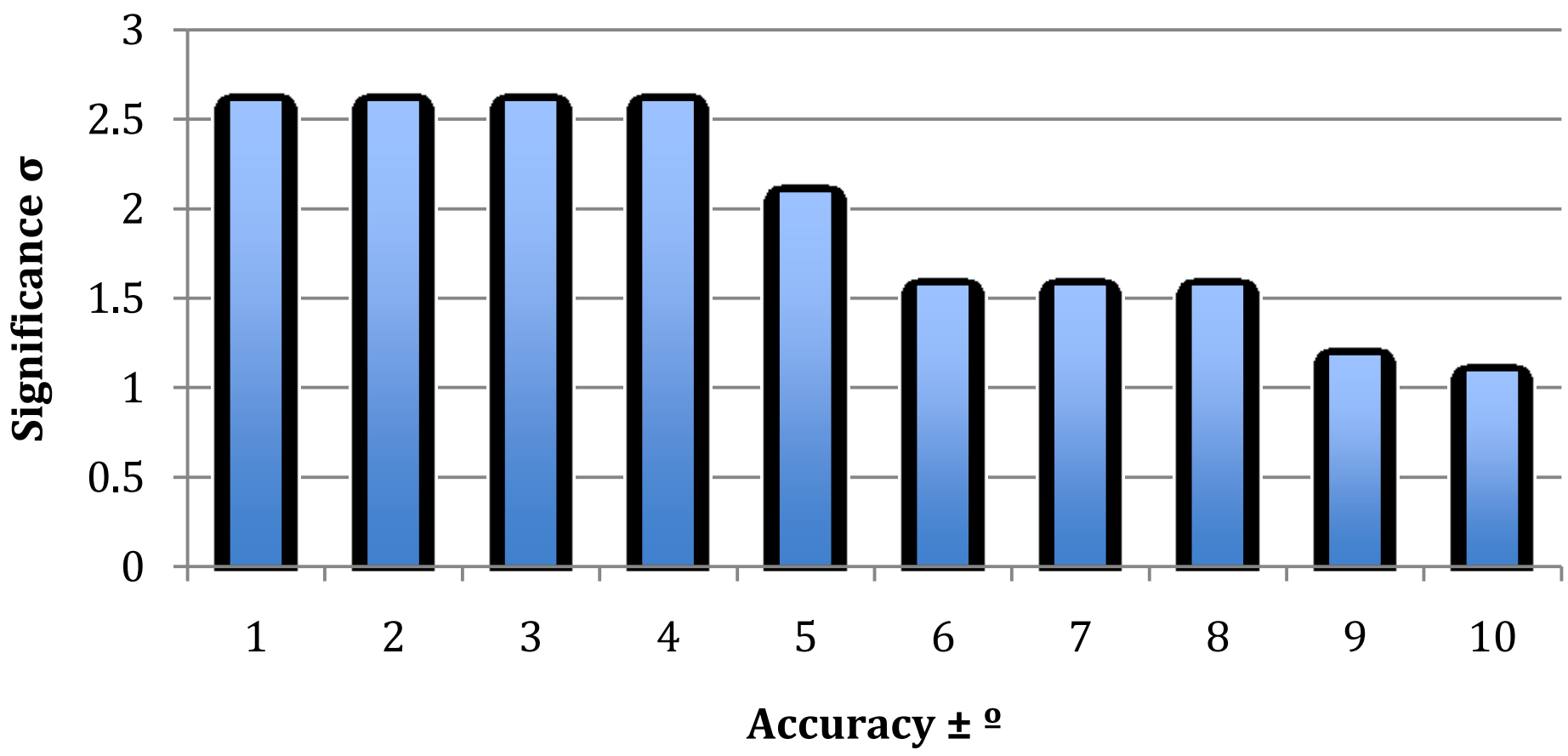

| Presumed Accuracy | 1º | 2º | 3º | 4º | 5º | 6º | 7º | 8º | 9º | 10º |
|---|---|---|---|---|---|---|---|---|---|---|
| Number of matches | 2 | 2 | 2 | 2 | 3 | 6 | 6 | 6 | 8 | 9 |
| Distance of result from expected value (σ) | 2.61 | 2.61 | 2.61 | 2.61 | 2.10 | 1.58 | 1.58 | 1.58 | 1.19 | 1.10 |

## Megara Hyblaea

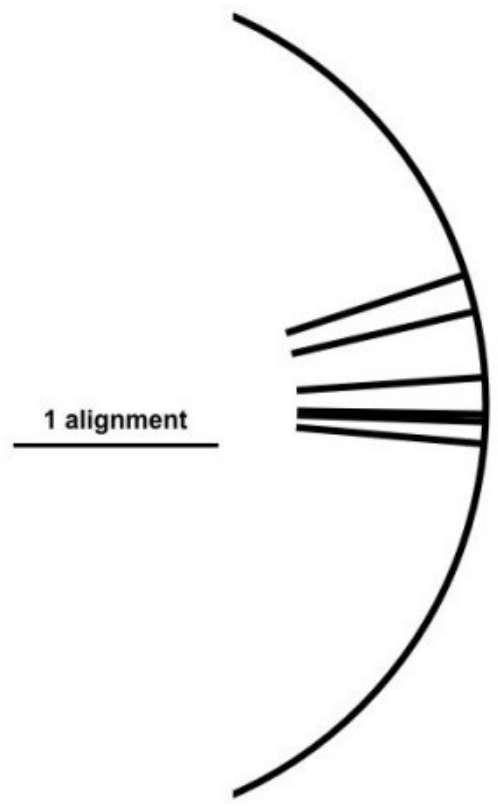

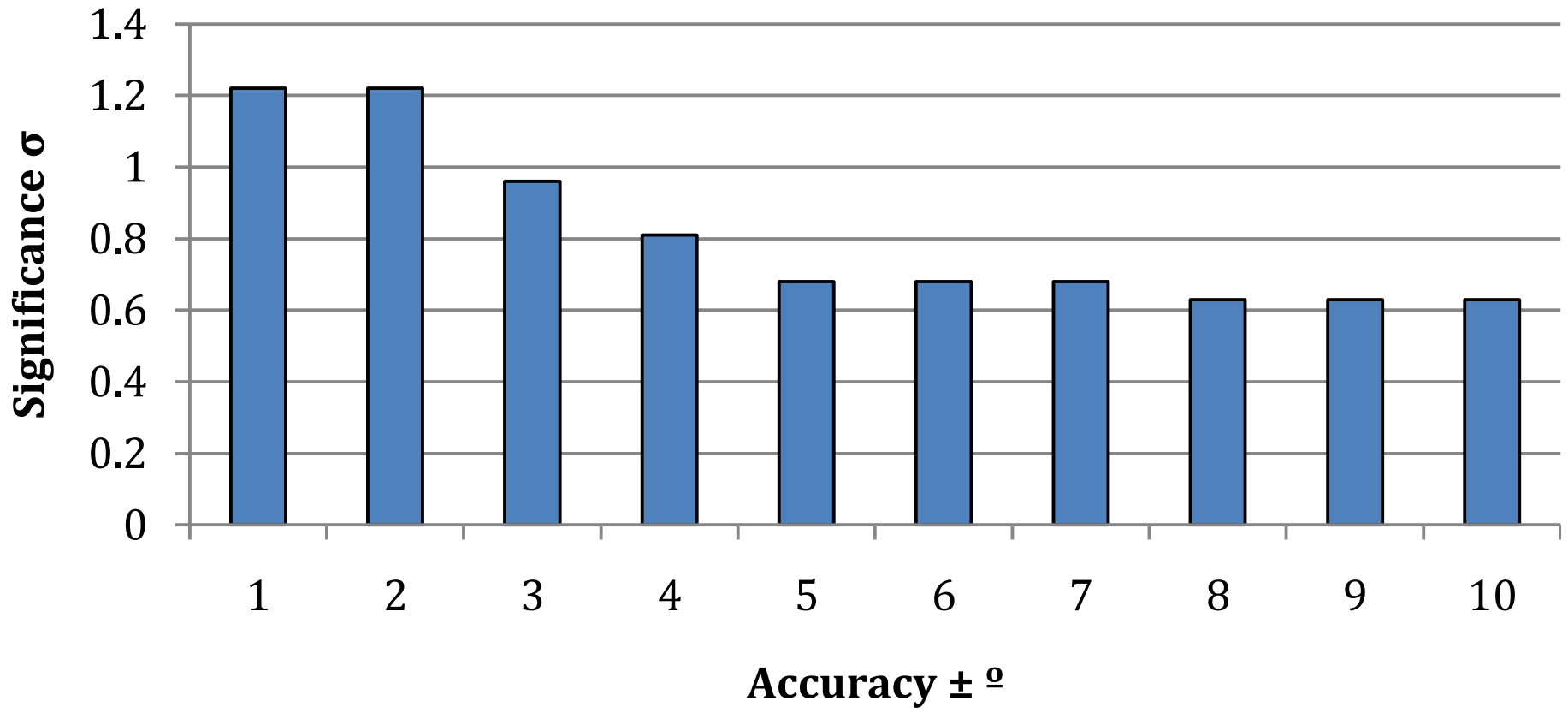

| Presumed Accuracy | 1º | 2º | 3º | 4º | 5º | 6º | 7º | 8º | 9º | 10º |
|---|---|---|---|---|---|---|---|---|---|---|
| Number of matches | 13 | 13 | 17 | 20 | 23 | 23 | 23 | 24 | 24 | 24 |
| Distance of result from expected value (σ) | 1.22 | 1.22 | 0.96 | 0.81 | 0.68 | 0.68 | 0.68 | 0.63 | 0.63 | 0.63 |

## Naxos

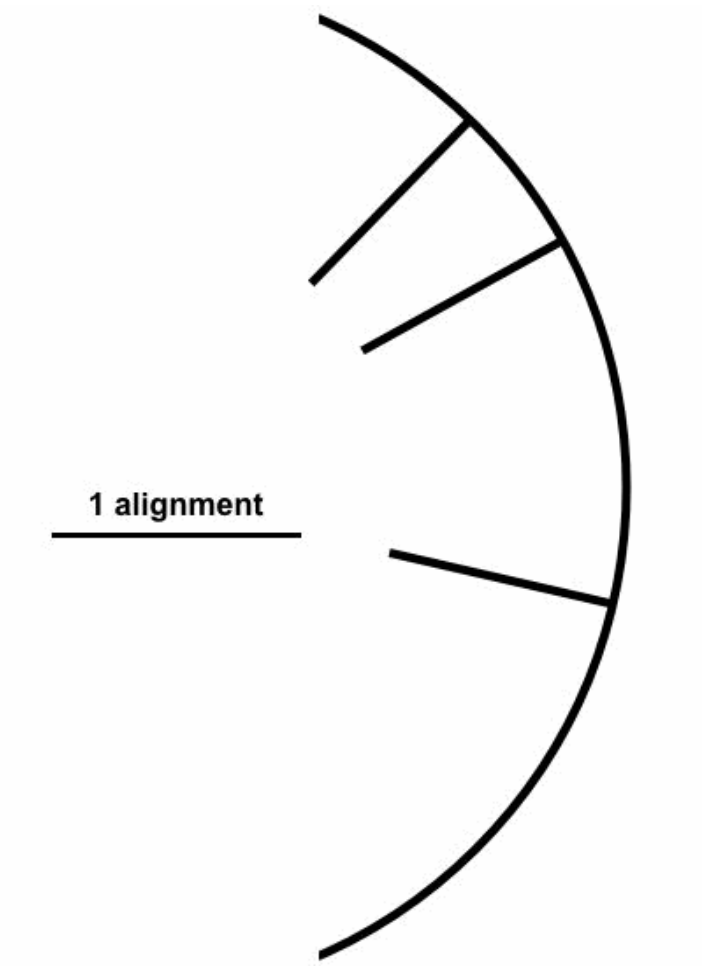

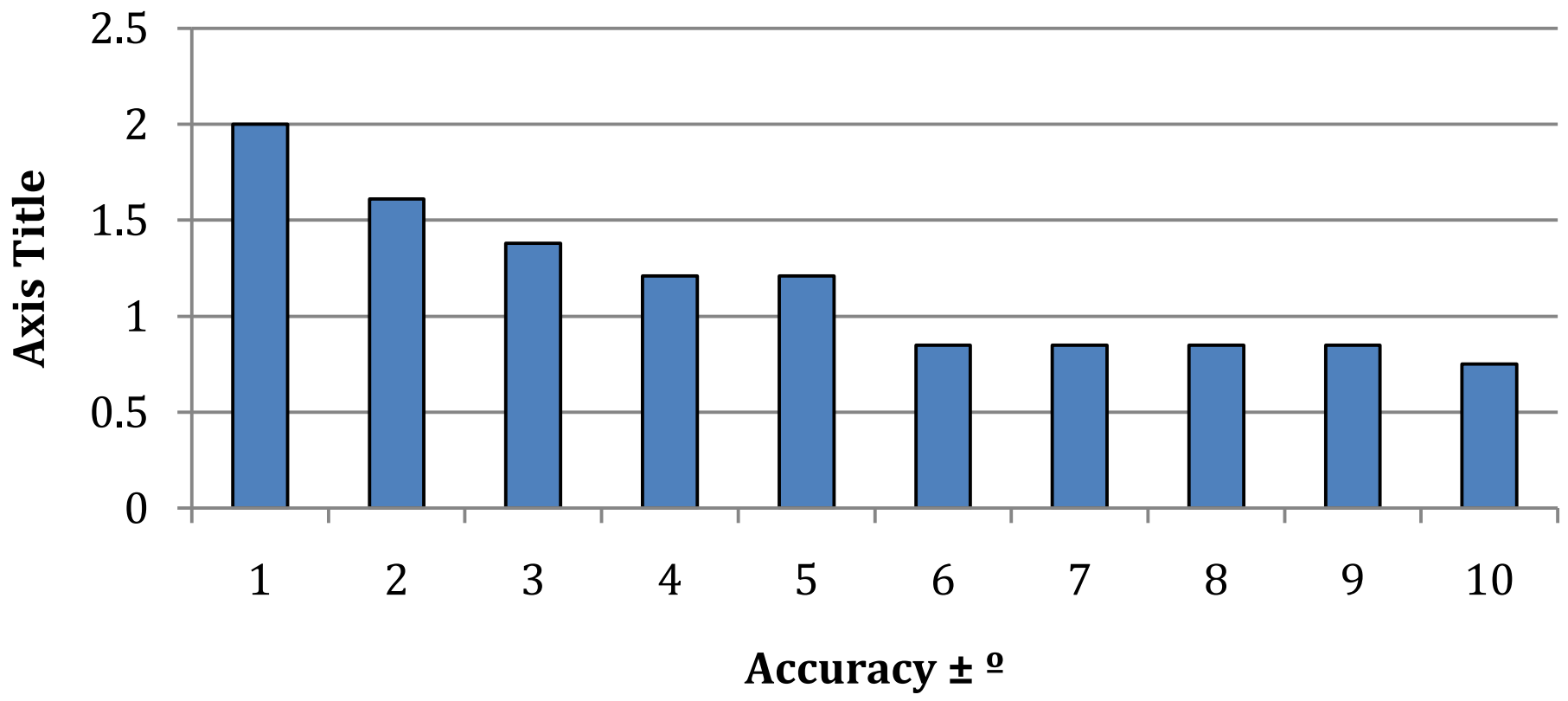

| Presumed Accuracy | 1º | 2º | 3º | 4º | 5º | 6º | 7º | 8º | 9º | 10º |
|---|---|---|---|---|---|---|---|---|---|---|
| Number of matches | 2 | 3 | 4 | 5 | 5 | 9 | 9 | 9 | 9 | 11 |
| Distance of result from expected value (σ) | 2.00 | 1.61 | 1.38 | 1.21 | 1.21 | 0.85 | 0.85 | 0.85 | 0.85 | 0.75 |

## Selinous

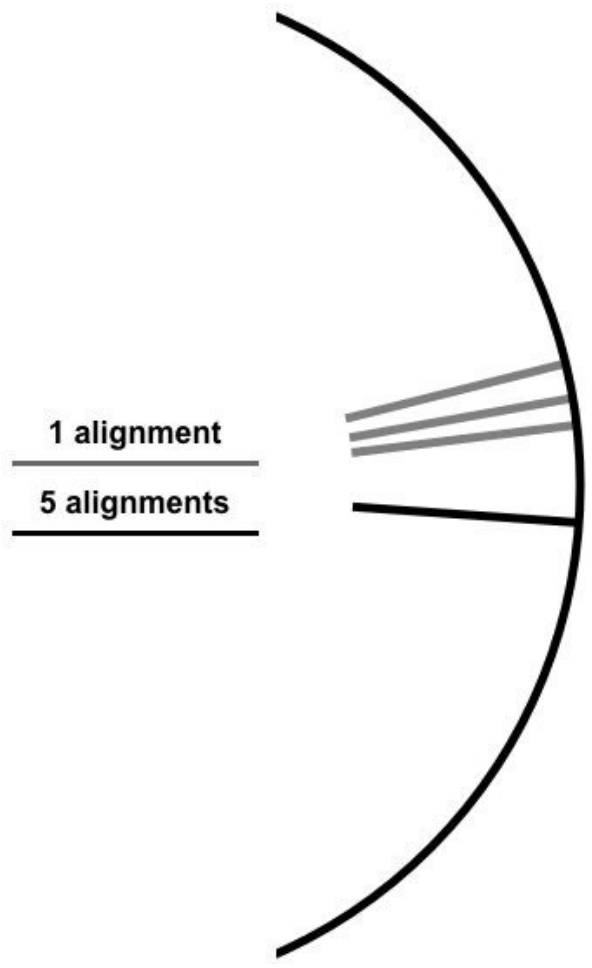

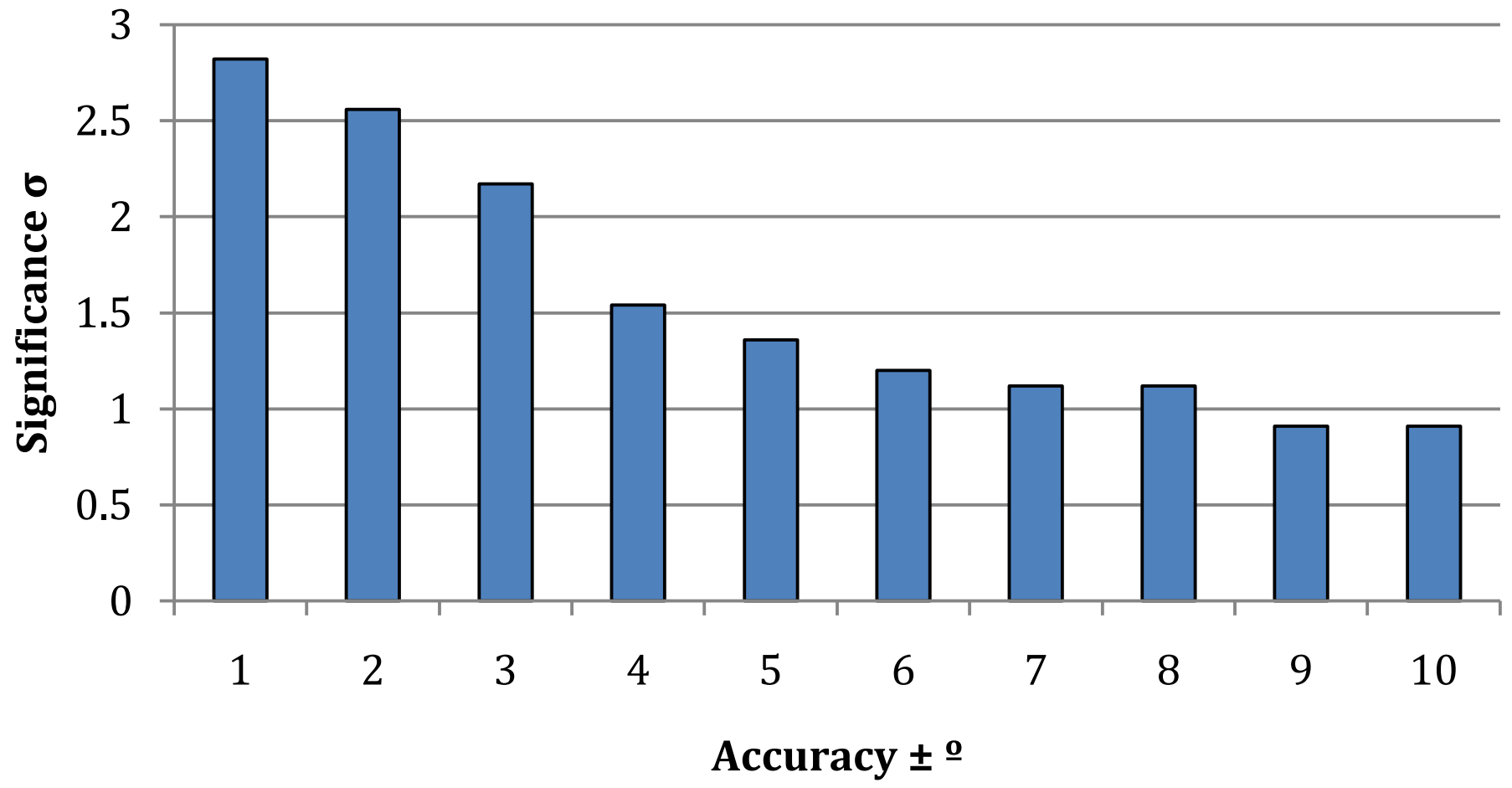

| Presumed Accuracy | 1º | 2º | 3º | 4º | 5º | 6º | 7º | 8º | 9º | 10º |
|---|---|---|---|---|---|---|---|---|---|---|
| Number of matches | 6 | 7 | 9 | 14 | 16 | 18 | 19 | 19 | 22 | 22 |
| Distance of result from expected value (σ) | 2.82 | 2.56 | 2.17 | 1.54 | 1.36 | 1.20 | 1.12 | 1.12 | 0.91 | 0.91 |

## Syracuse

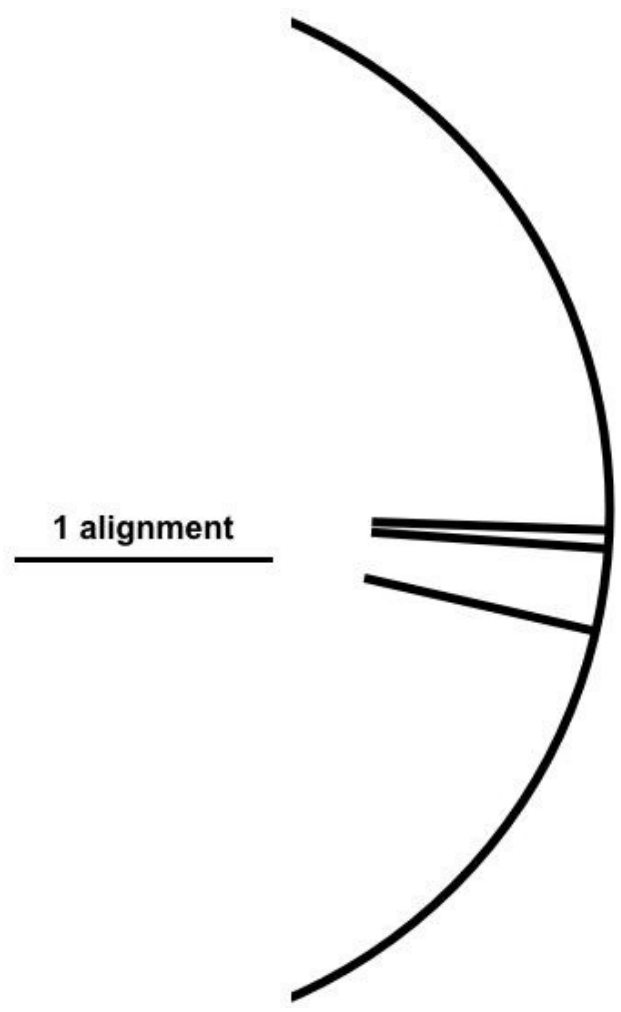

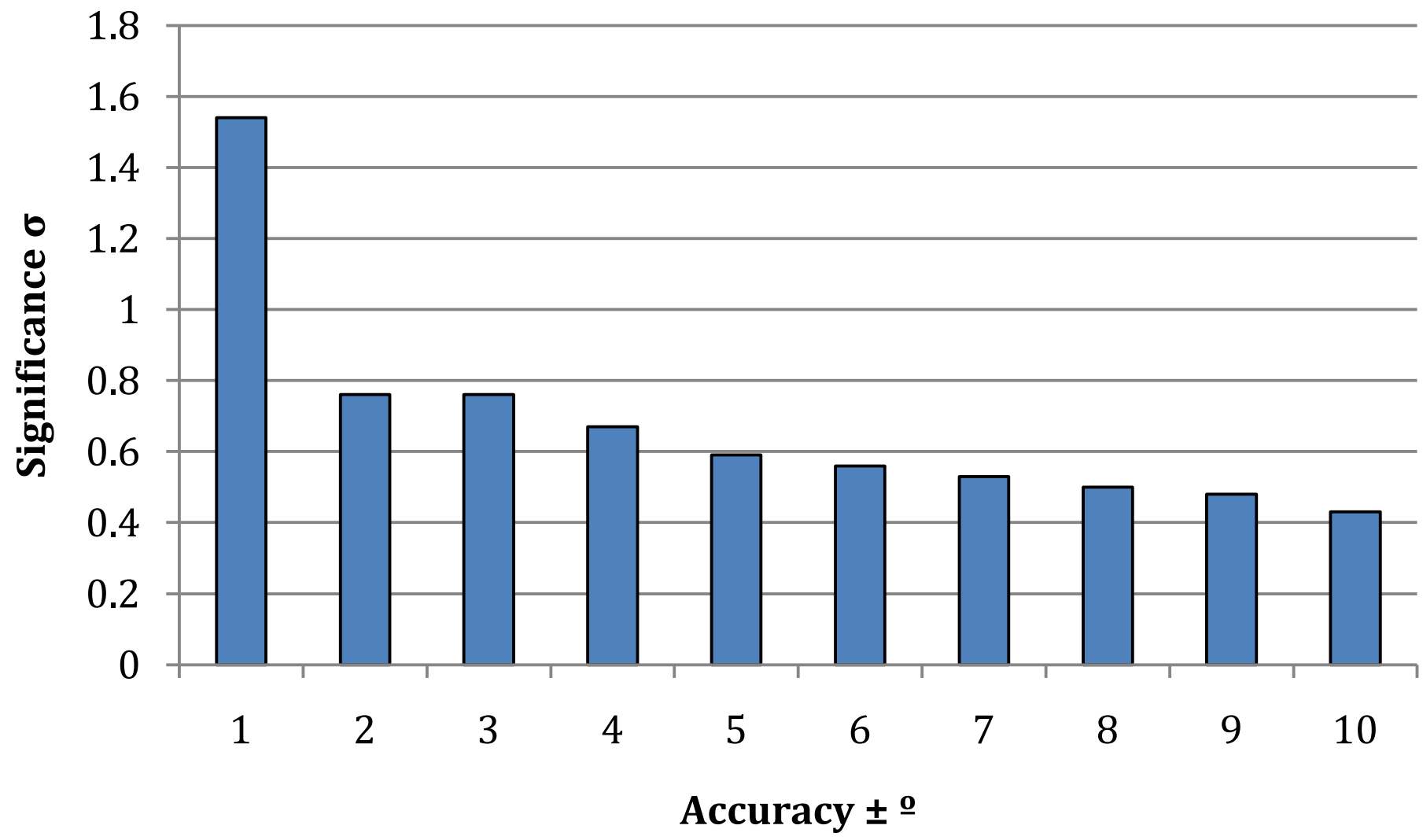

| Presumed Accuracy | 1º | 2º | 3º | 4º | 5º | 6º | 7º | 8º | 9º | 10º |
|---|---|---|---|---|---|---|---|---|---|---|
| Number of matches | 3 | 10 | 10 | 12 | 14 | 15 | 16 | 17 | 18 | 20 |
| Distance of result from expected value (σ) | 1.54 | 0.76 | 0.76 | 0.67 | 0.59 | 0.56 | 0.53 | 0.50 | 0.48 | 0.43 |

Some common features occur. The most obvious is that the quoted significance is positive in all cases. It should be emphasised that any figure below 2 cannot be said to be significant. The positive values are a result of the target alignments being known parts of the total data set. Another factor is that significance reduces as presumed accuracy lessens. This is also not surprising. This would be a reiteration of the widely accepted fact that the more certain you can be about an alignment the more likely it is you can say something about it. It is also for this reason that the lack of significance in any of the small samples, Gela, Himera and Naxos should expected. The sample sizes are simply not large enough to say anything statistically significant.

However there are striking differences between some samples. Some locations would seem to have distinctive patterns of alignments. Selinous would seem to have a distinctive pattern of alignment, if you accept that the Greeks planned to one degree of accuracy. Selinous is a site where this claim is particularly plausible as many of the temples are aligned with the street grid. This alignment is coherent over a distance of more than a than a kilometre. The early date of Temple E on site in the late seventh century BC (Gullini, 1985) would suggest this gridding was conceived by this period at the latest. The lack of any significant pattern at Megara Hyblaea would also seem to suggest that this is a genuinely distinctive alignment. However, the adherence to the grid plan could be consistent with a topographical rather than astronomical motivation. Himera too has what might be a significant preference for some alignments, but at this site too the temples are aligned to within a degree of the street grid, which is aligned to 67º.

It is the alignment of temples at Akragas that most strongly suggests that astronomy may have been a factor in the orientation of Greek temples. As with other sites there is a general preference for east. This need not be due to observational astronomy. Greek religion practice did not occur in the temples but rather at the altar, which was almost always in front of the temple (Yavis, 1949, p. 56). Therefore religious ritual would be performed in the open. There are therefore good mundane reasons why east would be the favoured side. Greek ritual was performed early in the morning. An altar in the west could be in the temple's shadow before sunrise. Not only would light arrive, but so too would the warmth. This would not just warm the people but the soil where they stood. An easterly orientation would also aid the evaporation of morning dew from the interior of the temple and Vitruvius (4.1) gives this reason for orientating libraries to the east. Looking at the locations of the temples at Akragas would suggest that this was not a primary factor in the orientation of the temples.

The temples of Akragas are of a similar or later date to those found at Selinous (De Miro, 1994; Marconi, 1929) and, broadly speaking, can be arranged into three groups. One cluster sits on top of the Acropolis overlooking the city to the south. The temples here, Athena Lindioi, Zeus Atabyrion and a temple of Demeter, that was not part of the survey all, face what was probably a distant horizon to the southeast. The buildings of modern Agrigento make it hard to be certain. A second cluster set of temples sit on the ridge that marks the south of the city and these seem to be arranged to face distant horizons rather than the next easterly temple along the ridge. The final group is the temple between these two regions of high ground.

The temple of Zeus Olympios stands next to the ridge and its alignment is skewed to the north to avoid facing the ridge. Further to the west lie the temples in the sanctuary of the chthonic deities. These all look up the valley to the distant horizon. Motivations regarding heat and light would be little affected by local topography. A rise of 4º would bring a horizon closer, but would only delay the arrival of the sun by around a quarter of an hour. A sun rising over a near horizon is just as warm as a sun over a far horizon. What is altered by a close horizon is the apparent position of the sun. The sun may not appear at the same time for all people present at an event. A distant horizon would mean that all people could share a collective observing experience. If the arrival of the Sun is part of a ritual then this removes a source of friction over the appropriate time to perform an action.

The results from Megara Hyblaea may seem to refute this argument. The results from Megara Hyblaea are only 1.22$\sigma$ away at most in comparison to 3.27$\sigma$ at Akragas, but Megara Hyblaea has a very different topography. It lies on the east coast of Sicily and from the city all that lies to the east is sea. Therefore the citizens had no topographical restrictions if they wished to point a temple at a distant sunrise.

It is possible that the orientations could be to star-rises rather than the Sun. As almost all the temples face a span of orientations within the range of the rising Sun, it is less problematic to propose that it was a solar event that was being observed.

## Genealogies of colonies

Trümpy (1997) has extensively researched the relationships between the calendars of the Greek cities. They were usually named after a major festival in the month and show similarities between related cities. The relationship between cities tracing their origins back to Corinth has led Freeth *et al.* (2008) to suggest that the Antikythera Mechanism is possibly a Syracusan device based on the similarity of the *parapegma,* or calendar, on the device to that of Tauromenion, which was a daughter city of Syracuse. Religion in ancient Greece was less about a personal relationship with a God, and more about the act of being a citizen. Citizenship was usually required to take part in the religious events of a city and part of the duties in helping the community maintain an amicable relationship with the gods. Therefore there is a sound reason why calendars related to religious events should be also guides to political allegiances. The citizens of a colony would have argued their ancestors where originally citizens of the mother city and so a similarity of the mechanics of appeasing the gods can be expected. However, this requires textual evidence. Most calendars from Greek cities are fragmentary if they exist at all. Is there a pattern in the alignment of Greek temples that would also connect genealogies of cities?

For Sicily there are three prominent genealogies in the dataset. The easiest genealogy to identify is Megara Hyblaea. The city name comes in part from Megara in the homeland. It is said that the inhabitants of Megara Hyblaea sent for an *oikist*, a founder, for their new city in the west of Sicily, Selinous (Thucydides 6.4.4). These two are analysed below as the Megarian colonies. Another early settlement in Sicily was Syracuse. Syracuse was said to have been founded from Corinth (Thucydides 6.3.2). In turn Syracuse founded Akrai, Helorus, Casmenae and Camarina. There are no obvious temples at Casmenae, which is why it doesn't appear in the data set. Syracuse, Akrai and Helorus are analysed below as the Corinthian colonies. Camarina is excluded from this set, as the temple would be better examined as a Rhodian construction for the reason below.

In the traditional histories Rhodes was comparatively late arrival to Sicily, founding Gela in 688 BC with the aid of Cretans. Gela is said to have founded Akragas around 580 BC (Thucydides 6.4.4). It also settled Camarina twice after the Syracusans had abandoned it (Pelagatti, 1966). The temple at Camarina dates from the mid-fifth century *after* the Geloan settlement. While the city grid and early fortifications may be Syracusan in origin, the temple would appear to be a product of Geloan settlers and so the temples of Akragas, Gela and Camarina are collected as the Rhodian lineage below.

## The Corinthian temples

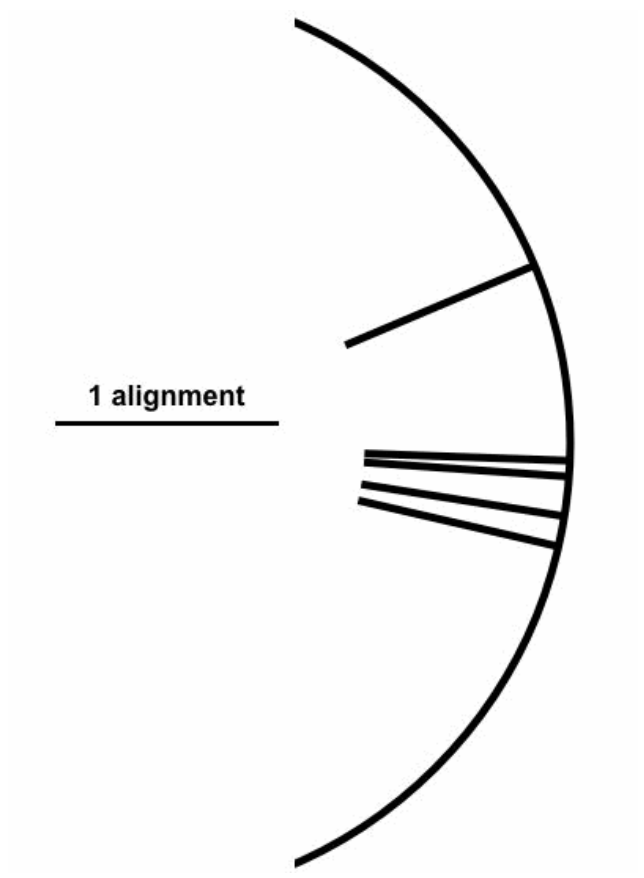

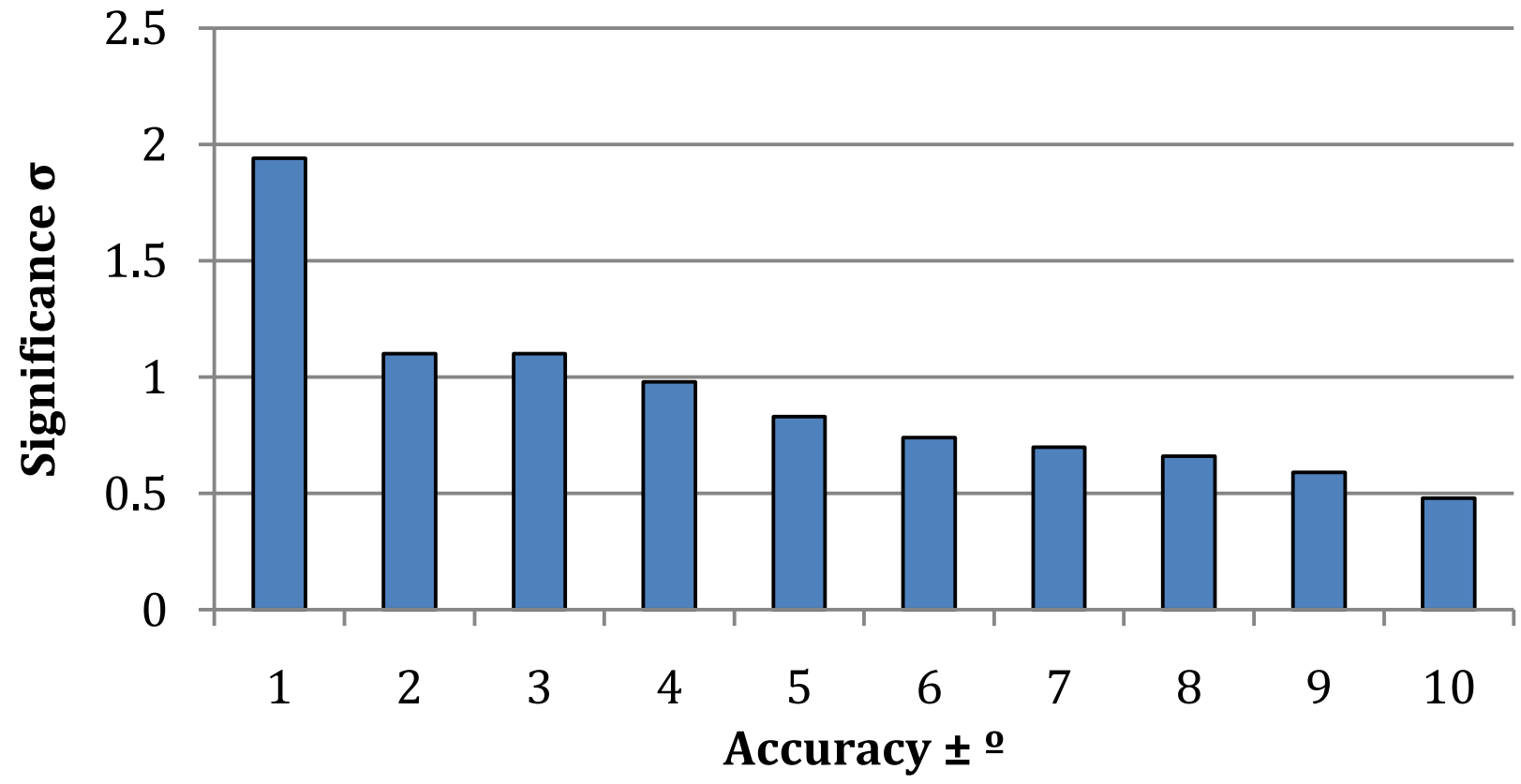

| Presumed Accuracy | 1º | 2º | 3º | 4º | 5º | 6º | 7º | 8º | 9º | 10º |
|---|---|---|---|---|---|---|---|---|---|---|
| Number of matches | 5 | 12 | 12 | 14 | 17 | 19 | 20 | 21 | 23 | 26 |
| Distance of result from expected value (σ) | 1.94 | 1.10 | 1.10 | 0.98 | 0.83 | 0.74 | 0.70 | 0.66 | 0.59 | 0.48 |

## The Megarian temples

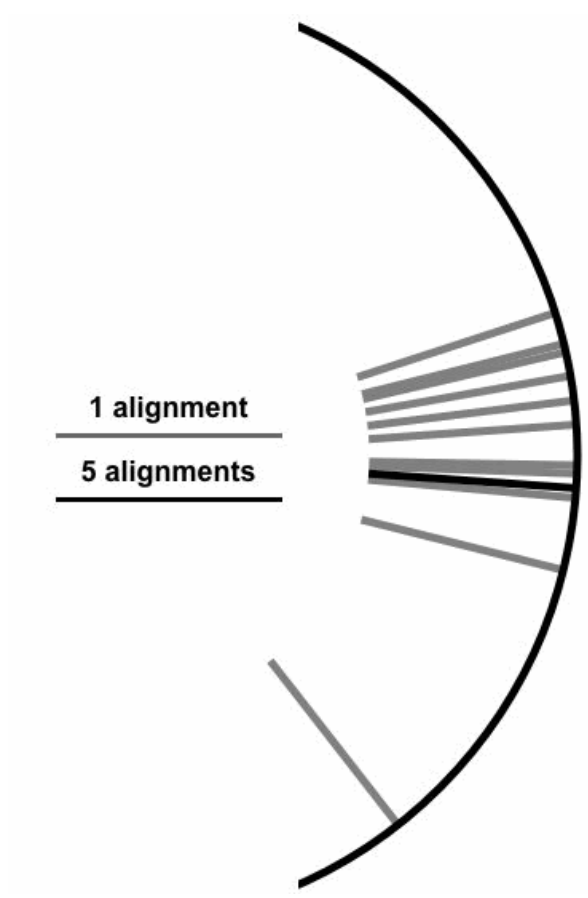

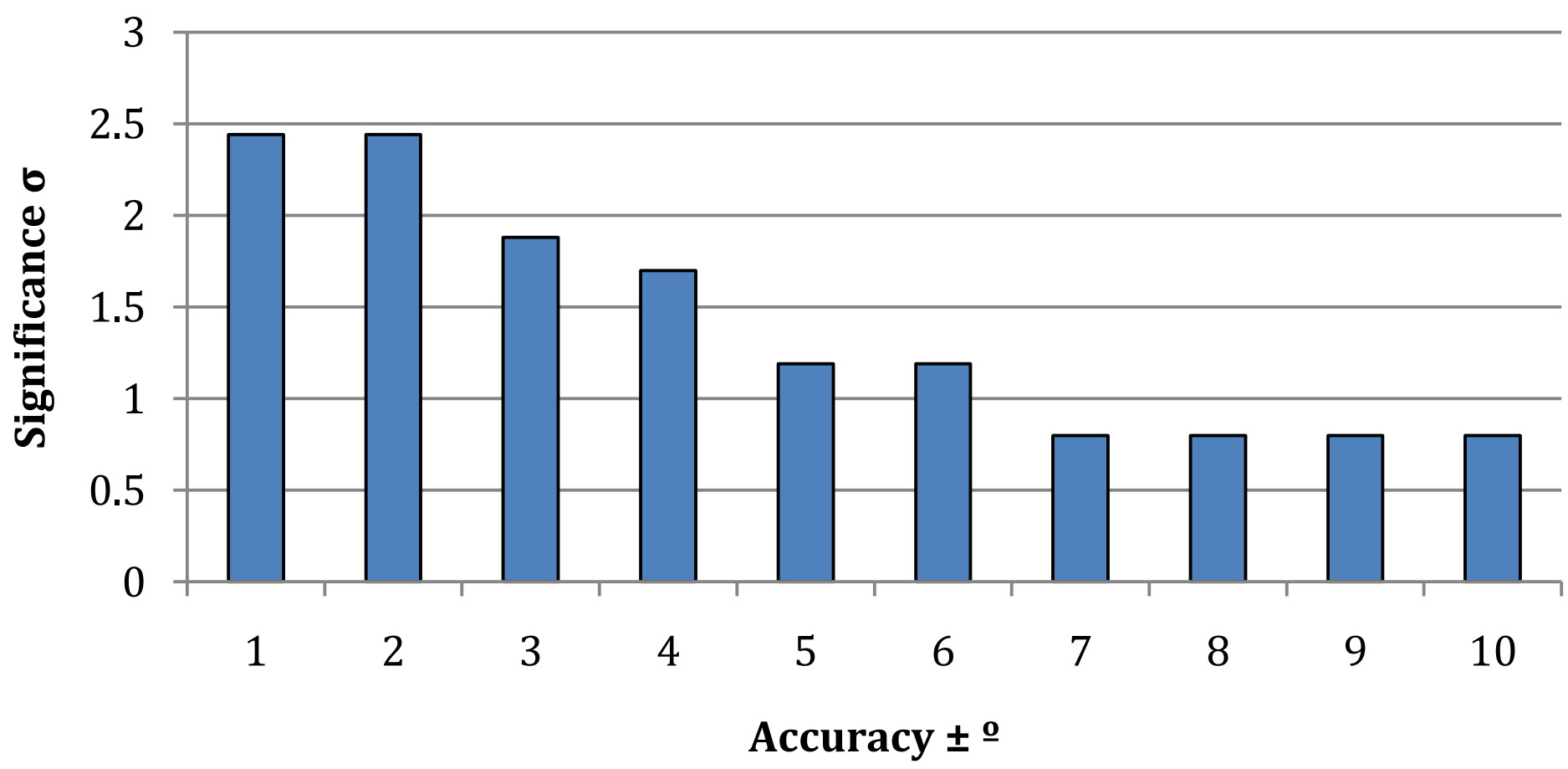

| Presumed Accuracy | 1º | 2º | 3º | 4º | 5º | 6º | 7º | 8º | 9º | 10º |
|---|---|---|---|---|---|---|---|---|---|---|
| Number of matches | 13 | 13 | 16 | 17 | 20 | 20 | 22 | 22 | 22 | 22 |
| Distance of result from expected value (σ) | 2.44 | 2.44 | 1.88 | 1.70 | 1.19 | 1.19 | 0.80 | 0.80 | 0.80 | 0.80 |

### The Rhodian temples

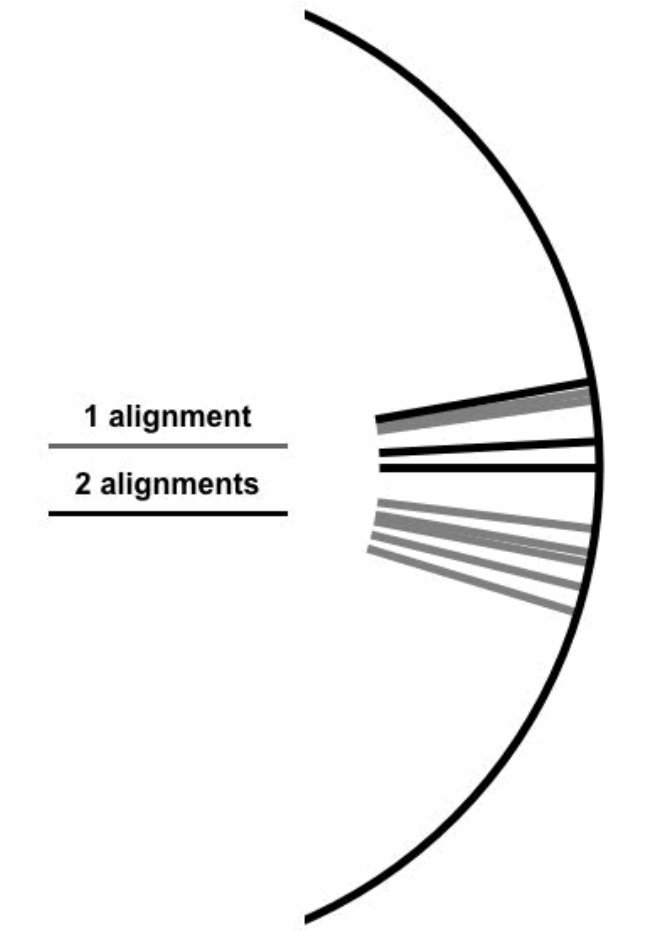

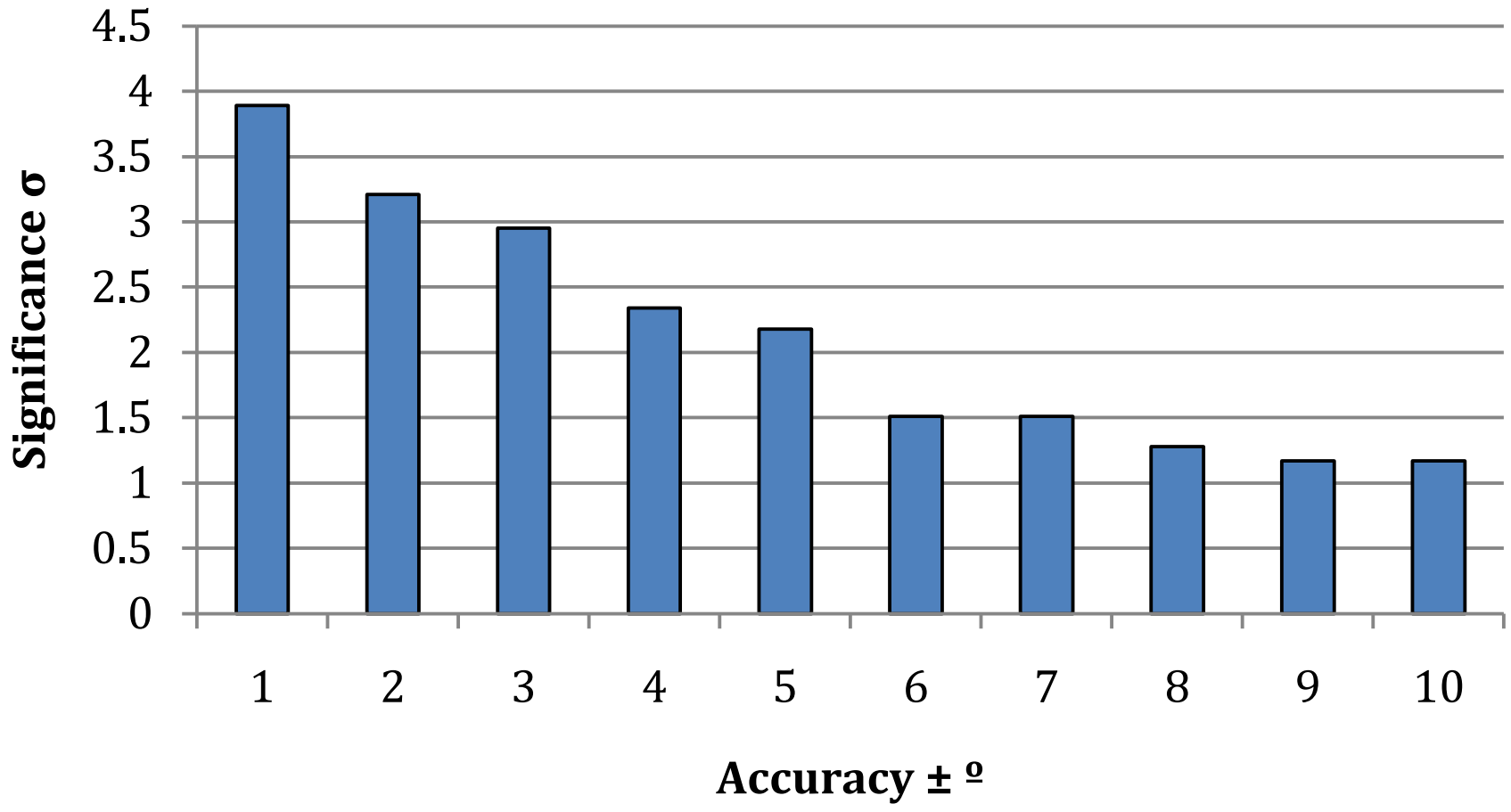

| Presumed Accuracy | 1º | 2º | 3º | 4º | 5º | 6º | 7º | 8º | 9º | 10º |
|---|---|---|---|---|---|---|---|---|---|---|
| Number of matches | 6 | 8 | 9 | 12 | 13 | 18 | 18 | 29 | 21 | 21 |
| Distance of result from expected value (σ) | 3.89 | 3.21 | 2.95 | 2.34 | 2.18 | 1.51 | 1.51 | 1.28 | 1.17 | 1.17 |

The above results are somewhat mixed. There is clearly not a statistical case for common Corinthian alignments in Sicily. The results are not significant, no matter what degree of accuracy you argue for in alignment. Superficially the Megarian colonies would appear to be distinctive. If an intention of high-accuracy is assumed, and the grid at Selinous would suggest it could, then there seem to be distinctively Megarian alignments. However, this is almost certainly the result of Selinous having a distinctive pattern of alignment with so many temples facing 96º. The orientations at Selinous were 2.82σ away from the expected result, and when Megara Hyblaea's temples are included this value drops.

In contrast the Rhodian colonies do seem to be distinctive if intentionally high-accuracy in their orientation can be assumed. Not only are they distinctive, but also collectively this value is higher than their individual analyses. This would suggest they are less distinctive when examine at a *polis* level as they are then compared with sister cities with which they share deliberate architectural similarities. The result is not so emphatic that the relationship is unquestionably real, but it is high enough to suggest that further investigation is required. One way of tackling this would be to examine the temples when grouped by their patron deity.

### Analysis of temples by patron deity

Herodotus argued that the Greek were unified by a common tongue, blood, religion and customs. It may be hard to strongly argue for a common religion. There was certainly a generic Greek religion, which local cults tended to be based upon. There were some Panhellenic sanctuaries, most notably at Delphi and Olympia. However because of the relationship between religious activity and political belonging in a *polis*, the gods worshipped in a *polis* were usually local forms of gods. These could be subtly different or grossly different. Zeus Olympios at Akragas was a celestial god. Zeus Melikhios at Selinous was probably a chthonic god, as in the form Melikhios was associated with Demeter and her quest to recover Persephone from the underworld. Therefore simply analysing temples by god may not be a sensible idea. The data exists to analyse temples to Aphrodite, Zeus, Demeter and Athena. However, even if an apparently significant pattern were to be found, the reader could question whether or not the temples are a culturally meaningful assemblage. For example are temples to Aphrodite Ourania in her heavenly aspect really comparable to temples of Aphrodite Pandemos? Further the lack of information about these temples means that in some cases we cannot be sure which aspects of the god are being celebrated. This problem is particularly striking in the case of the temples of Athena, which provide the best historical evidence. The temples of Athena are so puzzling they will be examined last as they require discussion at some length.

## Aphrodite

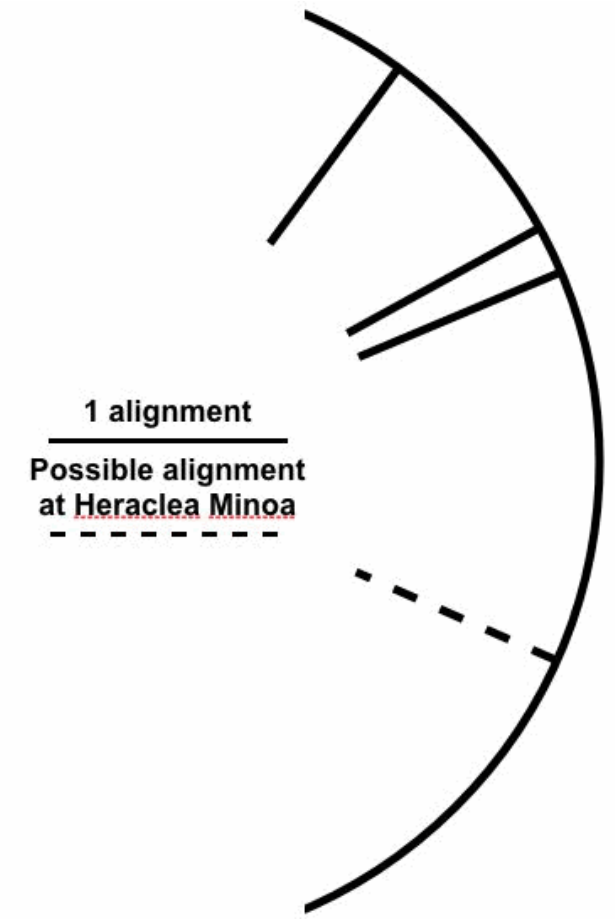

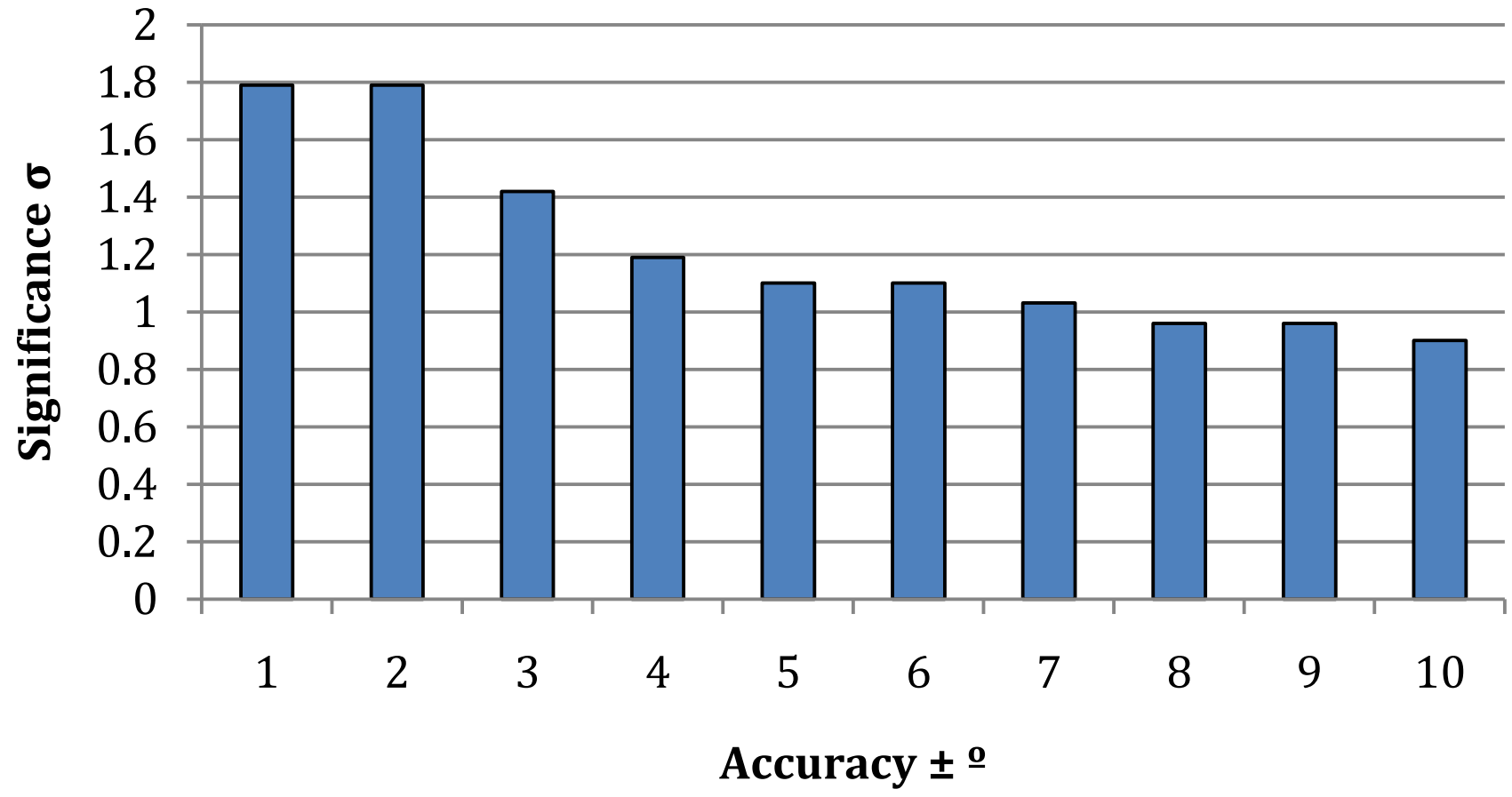

| Presumed Accuracy | 1º | 2º | 3º | 4º | 5º | 6º | 7º | 8º | 9º | 10º |
|---|---|---|---|---|---|---|---|---|---|---|
| Number of matches | 4 | 4 | 6 | 8 | 9 | 9 | 10 | 10 | 11 | 12 |
| Distance of result from expected value (σ) | 1.79 | 1.79 | 1.42 | 1.19 | 1.10 | 1.10 | 1.03 | 1.03 | 0.96 | 0.90 |

## Zeus

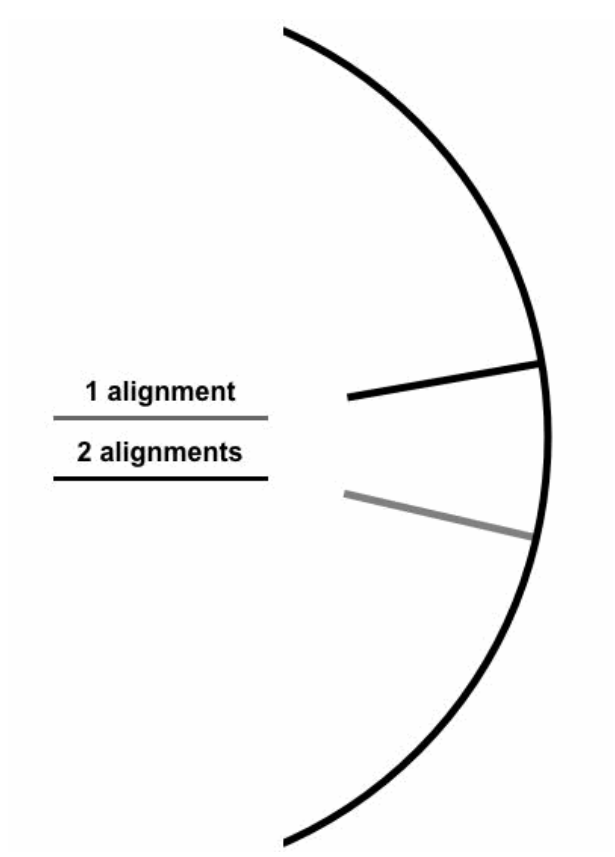

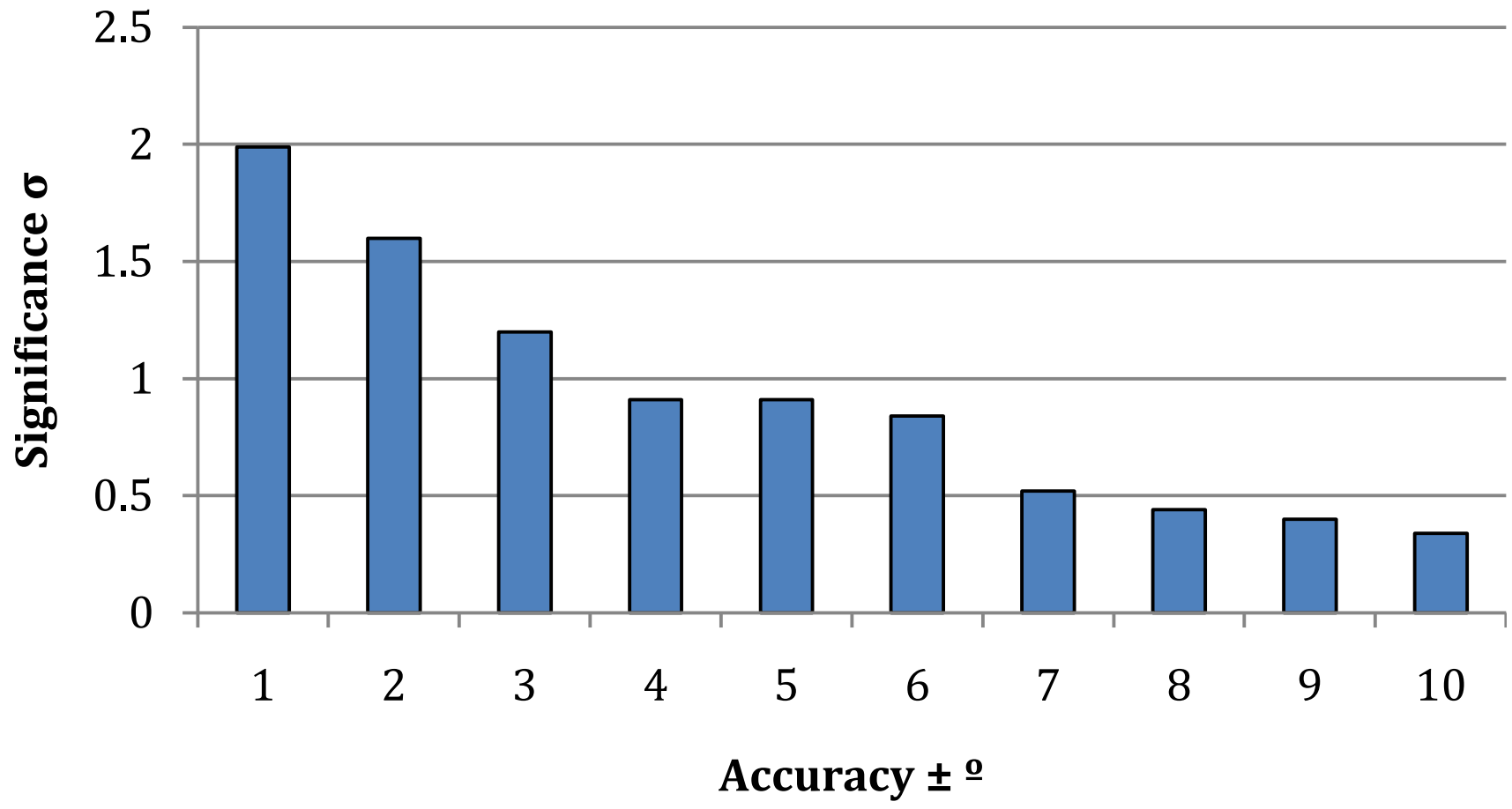

| Presumed Accuracy | 1º | 2º | 3º | 4º | 5º | 6º | 7º | 8º | 9º | 10º |
|---|---|---|---|---|---|---|---|---|---|---|
| Number of matches | 2 | 3 | 5 | 8 | 8 | 9 | 17 | 20 | 22 | 25 |
| Distance of result from expected value (σ) | 1.99 | 1.60 | 1.20 | 0.91 | 0.91 | 0.84 | 0.52 | 0.44 | 0.40 | 0.34 |

## Demeter

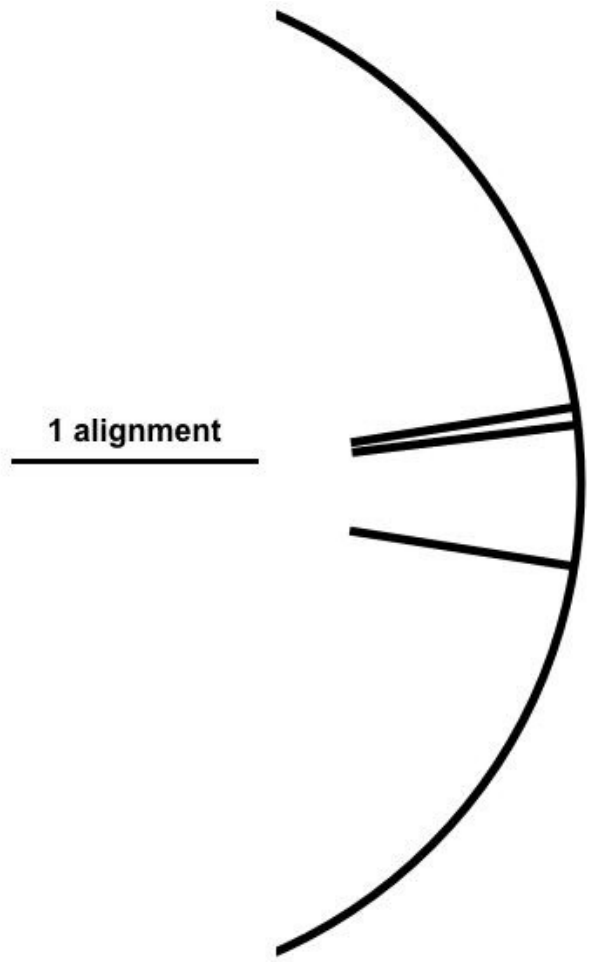

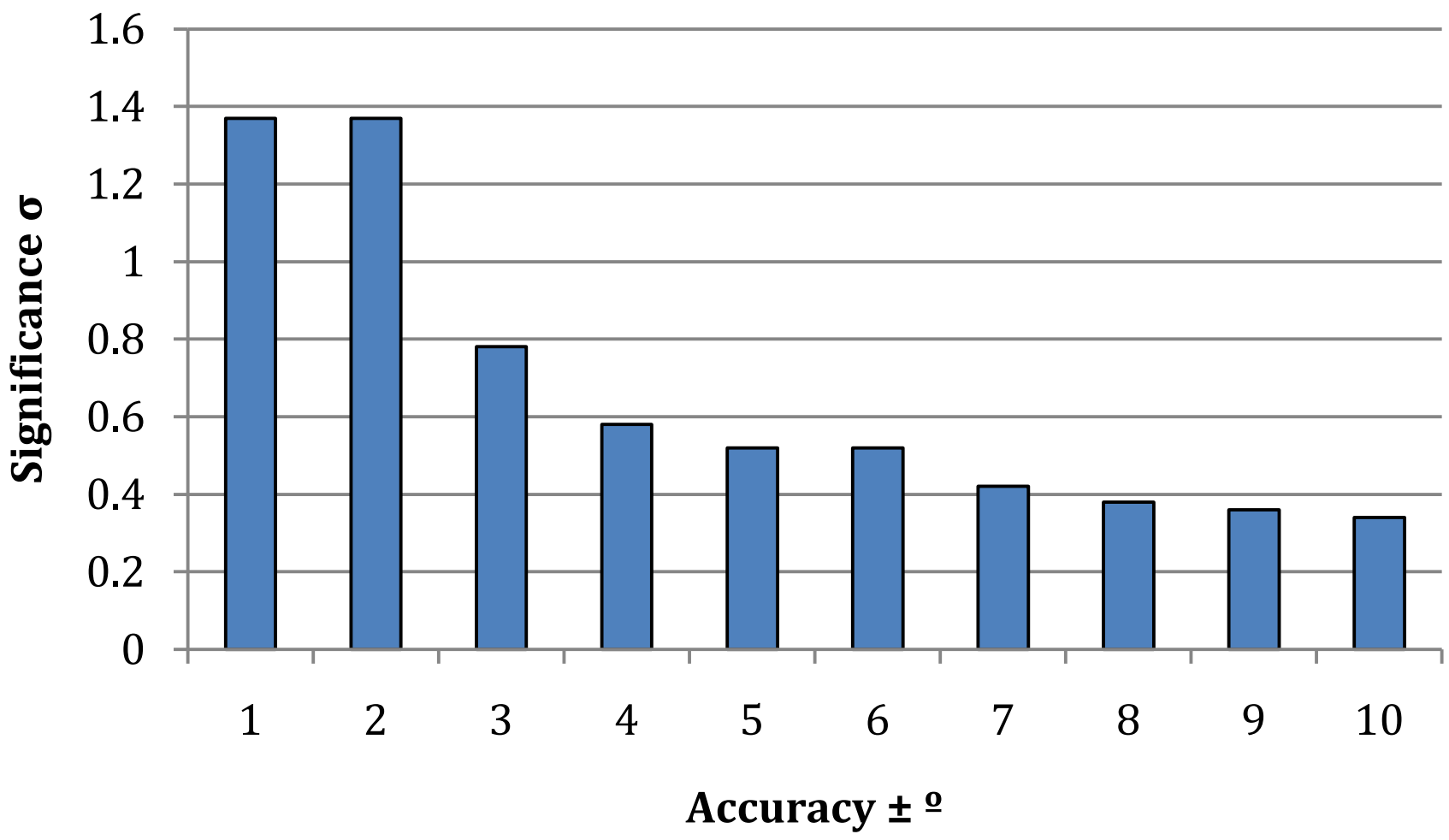

| Presumed Accuracy | 1º | 2º | 3º | 4º | 5º | 6º | 7º | 8º | 9º | 10º |
|---|---|---|---|---|---|---|---|---|---|---|
| Number of matches | 4 | 4 | 10 | 15 | 17 | 17 | 21 | 23 | 24 | 25 |
| Distance of result from expected value (σ) | 1.37 | 1.37 | 0.78 | 0.58 | 0.52 | 0.52 | 0.42 | 0.38 | 0.36 | 0.34 |

The temples at Herakleia Minoa complicate analysis of the Aphrodite temples. There are two temple-like buildings there. Diodorus Siculus tells us one is a temple of Aphrodite and the other is the tomb of Minos, who was said to have been murdered there while pursuing Daedalus. Unfortunately it is not possible to tell which is which. Given the preference for solar alignments it would seem reasonable to assume it is the most northerly orientated of the temples that is to the goddess, and that the second temple is a mistakenly identified tomb. This is not certain though. I have counted the northern temple as the temple of Aphrodite as it produces the less impressive results. The sample sizes for the three gods are small, only three or four temples and the aspects of the gods uncertain. Their alignments *might* have an astronomical significance, but the lack of data means this could only be demonstrated with substantial cultural evidence.

## The temples of Athena

There are far more temples of Athena to analyse. Three in Gela, and another one in each of Akragas, Camarina, Himera and Syracuse.

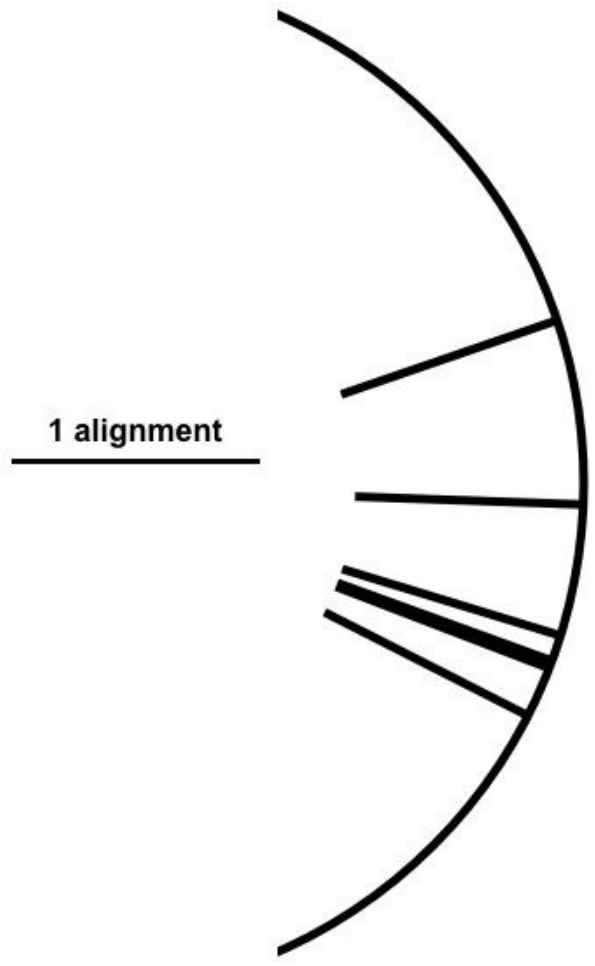

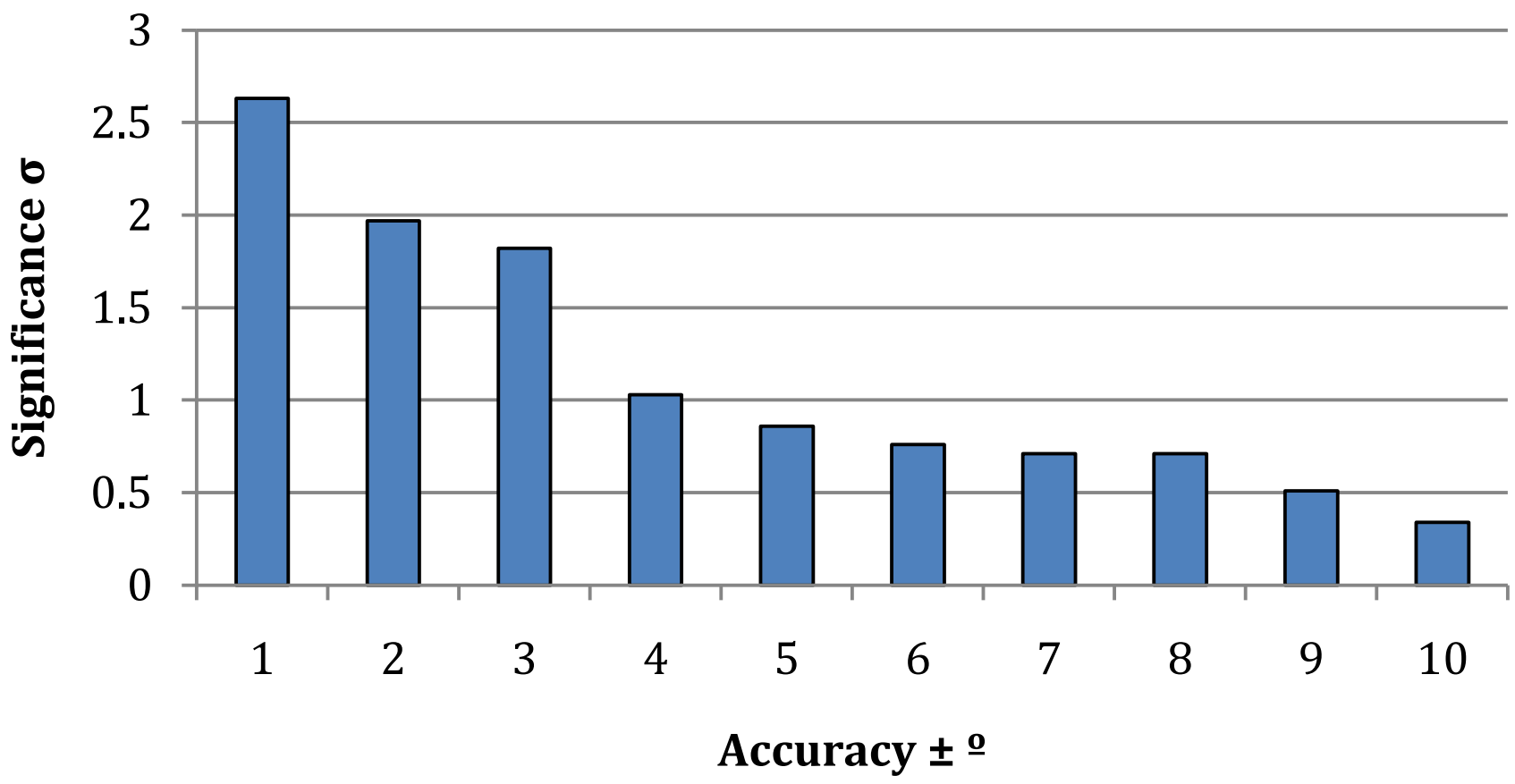

| Presumed Accuracy | 1º | 2º | 3º | 4º | 5º | 6º | 7º | 8º | 9º | 10º |
|---|---|---|---|---|---|---|---|---|---|---|
| Number of matches | 5 | 8 | 9 | 18 | 21 | 23 | 24 | 24 | 28 | 31 |
| Distance of result from expected value (σ) | 2.63 | 1.97 | 1.82 | 1.03 | 0.86 | 0.76 | 0.71 | 0.71 | 0.51 | 0.34 |

The temples of Athena might have a distinctive orientation if a high-accuracy is assumed. However, the results above would seem to be less impressive than the results for Rhodian colonies. As so many of the temples in this sample are Rhodian a better explanation would be one that combined both the Rhodian and religious elements of the temples.

As mentioned above deities are not uniform and have local aspects. In the case of Akragas, the temple of Athena is to Athena Lindioi, which sits overlooking the city. At the two other Rhodian cities, the temples of Athena sit on the high ground overlooking the cities. Examining solely the Rhodian Athena temples gives the result below.

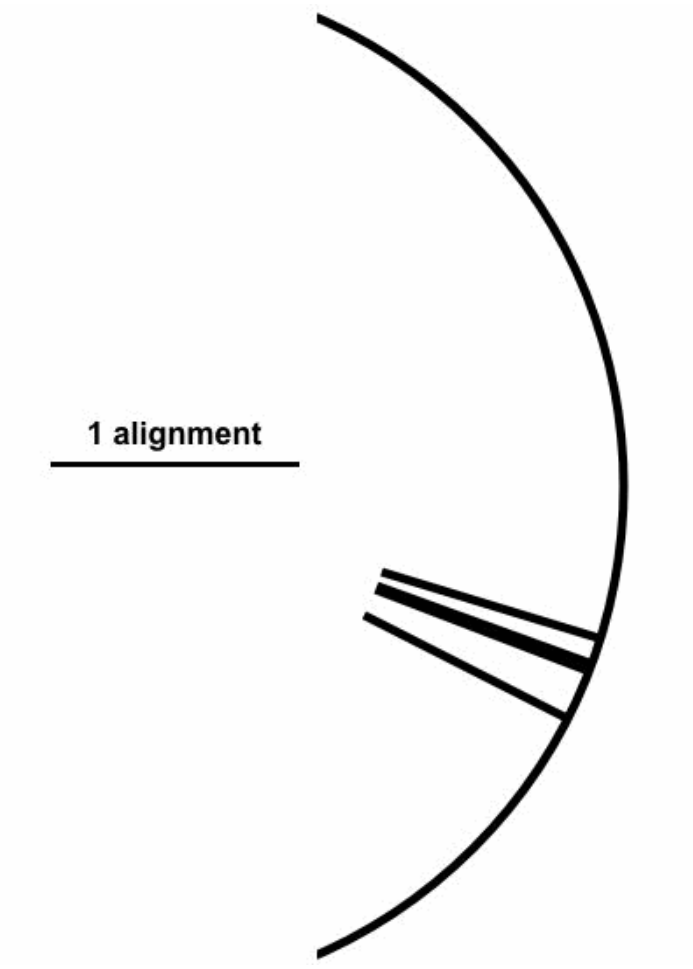

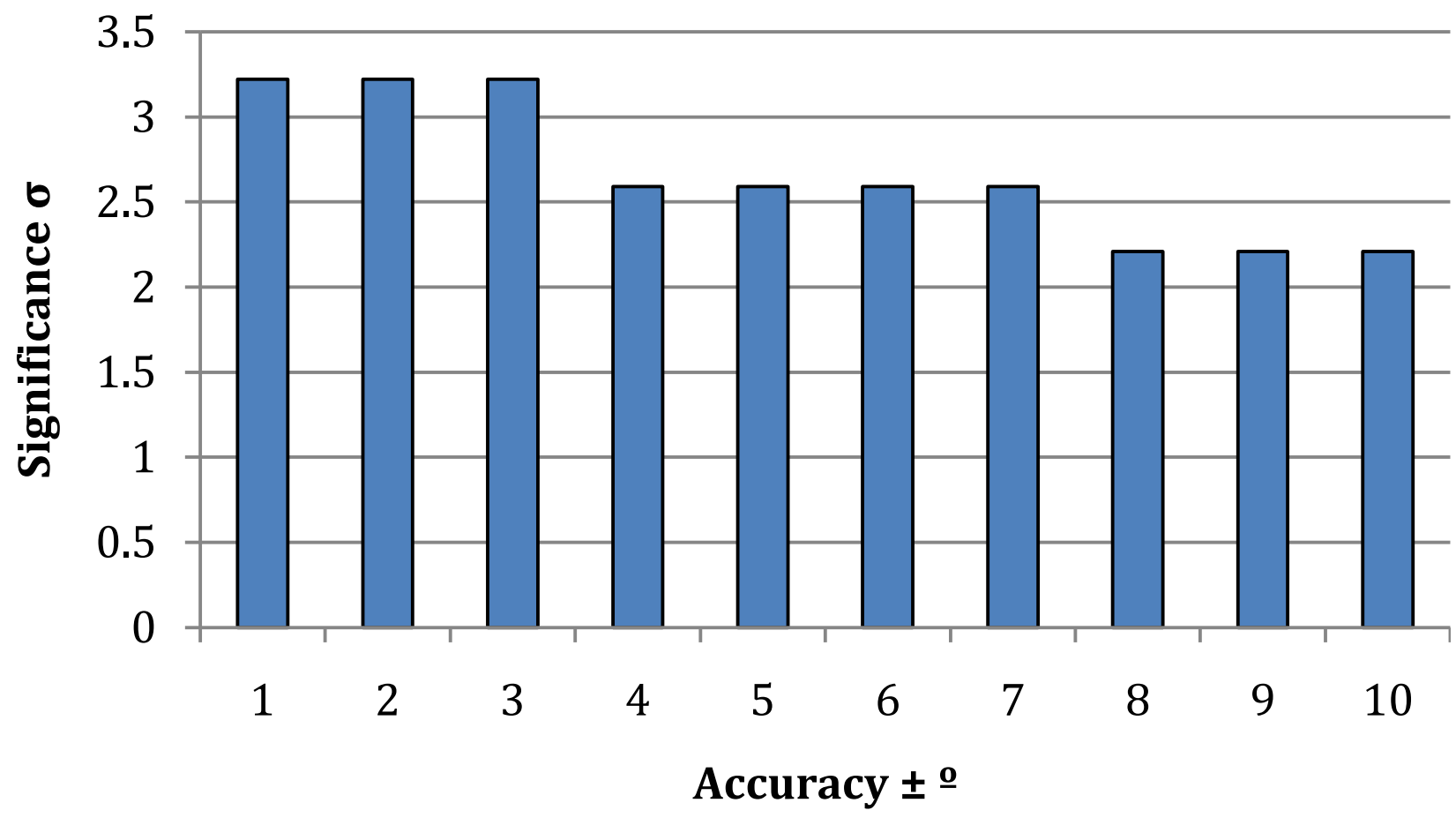

| Presumed Accuracy | 1º | 2º | 3º | 4º | 5º | 6º | 7º | 8º | 9º | 10º |
|---|---|---|---|---|---|---|---|---|---|---|
| Number of matches | 2 | 2 | 2 | 3 | 3 | 3 | 3 | 4 | 4 | 4 |
| Distance of result from expected value (σ) | 3.22 | 3.22 | 3.22 | 2.59 | 2.59 | 2.59 | 2.59 | 2.21 | 2.21 | 2.21 |

The figures are not as unusual as for the Rhodian lineage results at high accuracy, but they are consistently above 2σ right across to 10º. Temples of Athena at Rhodian colonies do seem to be pointing in a direction where few other temples face. Ideally I would now reveal that the temple of Athena at Lindos, Rhodes, also points to around 117º. Sadly a survey by the Danish Archaeological Institute has the Hellenistic temple pointing to 32º (Dyggve, 1960, p. 135) or 37.5º or 216º (Liritzis & Vassiliou, 2003, p. 95), which is too far to the north to face the rising sun.

Another objection which can be raised is that this result could be due to the three iterations of the Temple of Athena on the acropolis of Gela. It can be argued that the shape of the Acropolis is partly responsible for the orientation of the Geloan temples, and rather than five temples, the sample size should be three. The temple at Camarina would be evidence against such restrictions. The temple of Athena currently stands slightly short of the peak of the hill overlooking the ancient town. The temple is more or less at the top of the hill, so the rise in slope around the temple is slight, and may be due to geomorphological processes since antiquity rather than fully indicative ancient topography, but it does show that there was likely to have been a freedom of orientation at Gela, and the reason the successive temples face a similar direction is that, for whatever reason, that direction was the to place they wanted their temple to face.

Not only do the Rhodian temples form a coherent cultural unit, but there is are good grounds for examining the non-Rhodian temples as part of a different cultural unit environment. It can be shown that the temples of Athena at Himera and Syracuse were built in a very different historical context to the Rhodian temples at the Himera and Syracuse constructions can be linked to an historical event. Both temples were built after the Battle of Himera in 480BC.

### The post-Himera temples

480BC marks a watershed in the development of a Greek identity. In was in this year that Xerxes invaded the Greek mainland. The Greeks collectively fought against a barbarian (non-Greek) enemy for their independence. Fighting alongside the other cities cemented a notion of brotherhood. This raises difficulties for a Sicilian Greek identity, as the major cities of Greek Sicily were not fighting the Persians. In 480BC they were fighting the Carthaginians in Sicily. Victory over the Carthaginians was won at Himera, though the date is uncertain. Herodotus says the victory occurred the same day as the defeat of the Persian navy at Salamis (Herod. 7.166). Diodorus Siculus puts it at the same time as the battle of Thermopylae (Diod. 11.24.1). These two events were around a month apart. Feeney (2007, pp. 43-46, 51) has argued that the confusion has a narrative purpose. History in this period was not so much a matter of chronology, but a tale of how things came to be. Time has a narrative function, and by fighting the Carthaginians at the same time as the mainland Greeks were fighting the Persians, the Sicilian Greeks could show how they too were defending Greek homes from the barbarian. Following the battle of Himera Gelon, the ruler of Syracuse, dedicated a temple to Athena in her form of Athena Nike, goddess of Victory. He returned to Syracuse and on his arrival dedicated a temple to Athena there too.

So far, above, the idea, which has been tested, is that astronomy was primarily used in the practice of Greek religion at temple sites. The preference for distant horizons would suggest an interest in observations yet, apart from the temples of Athena at the Rhodian cities of Sicily, there is no strong evidence for astronomical practice at temples. Another suggestion is that astronomy was used in the construction of temples. If there was a general interest in an easterly orientation, it may be enough that the temple faced sunrise on the day that the plan was laid out. At most sites this would be impossible to test. The dating of most temples in Sicily is only to within a decade or two on stylistic grounds. However, the start of temple construction following the battle of Himera can be dated to the year.

In the case of the temple of Athena Nike at Himera the orientation is to 71º azimuth (Aveni & Romano, 2000, p. S54)(Marconi, 1931) and a declination of 15º. This means the temple would face sunrise on the 14th of August[1]. It is not certain if this would be plausibly close to the date of the battle of Thermopylae. Labarbe (1959) dates the battle to as early as the 31st of July. Dascalakis (1962, pp. 140-169) and Hignett (1963, pp. 448-449) both date the event to the 29th of August. Sacks (1976) rejects this date in favour of the 19th of September. This is due to an eclipse on the 2nd of October, which would suggest he is using a Julian calendar. The proposed eclipse is not strong evidence for the date of the battle as it was only visible as a partial eclipse from Greece, and is unlikely to have noticeably dimmed the sky (Mosshammer, 1981). The lack of concordance means that it would be futile to attempt to date the temple with respect to the Greek mainland, but it idea that it faces sunrise on the day its foundation could still be tested against the temple at Syracuse.

The Athenaion at Syracuse faces 92º and, if the horizon is flat, a declination of around -2º. It is hard to be certain as the ruins lie beneath the modern city (Orsi, 1919). This would be consistent with a date of around the 28th of September, which, in 480BC was also the day of a new moon and so possibly the start of a new Syracusan month. This date is be consistent with Gelon leaving Himera and arriving back home and founding a new temple. The return of a victorious army from Himera around 250km by foot through central Sicily, or a by longer but almost certainly faster trip by ship around the coast, would be feasible within this timescale. This may be plausible, but not convincing. The alignment of the temple of Himera to the north of east made a more southerly orientation quite likely anyway.

There is another post-Himera temple we can examine. The temple of Olympian Zeus was built by Theron of Akragas following victory at the battle (Diod. 11.26.2, 13.82.1-4). Akragas was closer to Himera, 110km through central Sicily. If Theron dedicated his temple on his arrival, then it should have an alignment between that of the temple of Nike at Himera and the Athenaion at Syracuse. The temple of Olympian Zeus faces an azimuth of 80º and a declination of +10º, which is sunrise around the 29th of August in the Gregorian calendar, and also the day of a New Moon in 480 BC. This allows around ten days to cross the terrain of central Sicily, if Theron's army was still at Himera when Gelon was dedicating his temple.

---

[1] This is a Gregorian date. The date in the Julian calendar is around 19th of August, which some planetarium software uses for BC dates.

The argument above cannot be used to date the battle of Himera, or other battles purely by astronomy. It does however give some circumstantial evidence in favour of the earlier dates and also is consistent with the alignment of temples towards the rising Sun when they are being built. This act of using astronomy in laying out a temple may also be consistent with the preference for distant horizons.

## A possible use of astronomy at Greek temples

If you use some simple geometry then laying out a rectangle, which would, form the base for a Greek temple is a trivial matter. One person could lay out an arbitrary baseline from which the rest of the gridding for a building is derived. Laying out a temple is not merely an architectural problem; it is also a religious matter. The value of an offering to the gods would be increased by the effort expended in making it. The building of a temple would also be a communal matter. A method, which allowed more people to work on a project, might be inefficient from an engineering point of view, but part of the dedication of the temple would be the process of making the temple.

A base line could be derived from laying out a line using ranging rods, rope and shadow from the Sun. If the horizon is distant then the area being worked upon is lit when the shadows are longest, and multiple work groups could lay out parallel lines by working at the same time. Building a temple is undoubtedly a religious event, which would explain why the alignment was to the east at the morning, rather than away from the west as the sun set. Aligning the temple as the sun set would be the more elegant solution from an engineering point of view. It would allow the shadows to be tracked as the sun fell in the sky, but it would be at the wrong time. If the alignment was a matter of laying out parallel lines for a grid to mark out the temple plan, then the fact that the alignment may not have exactly hit the sunrise point may be of no great concern. Such an approach to alignment would also explain why Selinous and Himera have topographically orientated grids; there was a better or more convenient local marker for alignment to 'east' than the Sun.

This does leave the alignment of the Rhodian Athena temples as a puzzle. It is possible that these temples were dedicated on a specific day of the year related to the cult of Athena Lindioi. In this case the shared alignments would be indicative of a shared religious cult. Comparative data is clearly needed but as the alignments of the post-Himera Athena temples show, the comparisons would have to be Athena in the same aspect as her cult in the Rhodian cities. It may be that the cultural evidence enabling the identification of similar cults elsewhere in the Mediterranean no longer exists.

## Conclusions

As argued in Salt (2009), the Greek temples of Sicily do seem to be orientated with respect to the Sun, but the precise relationship between temple alignment and the Sun remains uncertain. The alignments of temples with respect to the city grids of Selinous and Himera would indicate that high-accuracy alignment was practiced by the Greeks when building temples. The topographical orientations of the temples of Akragas also supports the idea of an interest in observation of distant horizons at these sites, and the results from the other Greek sites are consistent with this. However, only at the *poleis* tracing their ancestry back to Rhodes is there any hint of inherited practice in the alignment of temples. Paradoxically this may mean that the survey shows both that the alignment of temples towards the sun was emphatically not the result of chance – and that it was of little importance to Greek religious practice. One way out of this problem is to argue that astronomy was an important part of the process of building the temple and it is this process that leaves its mark in the alignment of temples. The alignments of the post-Himera temples are consistent, if not conclusive, with this proposal.

Greek temples are an attractive feature to examine for archaeoastronomy. They often have an unambiguous axis of alignment and frequently have an attribution allowing for the development of detailed hypotheses. There will doubtless be further studies of such temples. However, there is room for a reassessment of where religious activity took place in relation to the temple. The altars are almost always east of the temple, which would suggest that religious events took place on the sunny side of the temple. This may be true of Greek temples in regions where easterly orientations are less emphatic and it should be possible to re-examine data to see if practice occurred on the sunny side elsewhere.

The term 'sunny side' is intentionally vague, and a stronger definition will require a greater discussion of the practice of Greek astronomy. The method used above, based on the Binomial distribution can be adapted for this by adjusting the value of $p$ in light of cultural evidence. Indeed the method shows that cultural information can massively transform the confidence we can have in our interpretation of results. The Binomial distribution also allows for more flexibility in debating the meaning of proposed alignments. While I believe that alignment to within a degree is possible, the discussion is far from closed. By presenting the results in a way that allows many variations in terms of both presumed accuracy and what a significant result would be the reader is not compelled to accept all my proposals to make use of my findings.